\documentclass[%
reprint,
superscriptaddress,
 amsmath,amssymb,
 aps,
pra
]{revtex4-2}

\usepackage{tikz}
\usetikzlibrary{
    arrows.meta,
    decorations.pathmorphing,
    calc,
    positioning
}

\usepackage{graphicx}
\usepackage{dcolumn}
\usepackage{bm}
\usepackage{xcolor}
\usepackage{braket}

\usepackage[utf8]{inputenc}
\usepackage{booktabs}
\usepackage{array}
\usepackage{tabularx}
\usepackage{makecell}
\usepackage{hyperref}
\usepackage{svg}

\begin{document}

\title{
Modelling quantum measurement dynamics: from decoherence to redundancy with site-hopping indistinguishable particles}

\author{Katja Schneeweiss}
\email{katja.schneeweiss@tuwien.ac.at}
\affiliation{Institute for Theoretical Physics, TU Wien, Wiedner Hauptstra\ss e 8-10/136, 1040 Vienna, Austria}
\affiliation{Atominstitut, TU Wien, Stadionallee 2, 1020 Vienna, Austria}
\affiliation{Vienna Center for Quantum Science and Technology (VCQ), TU Wien, Vienna, Austria}

\author{Tom Rivlin}
\email{tom.rivlin@tuwien.ac.at}
\affiliation{Atominstitut, TU Wien, Stadionallee 2, 1020 Vienna, Austria}
\affiliation{Vienna Center for Quantum Science and Technology (VCQ), TU Wien, Vienna, Austria}

\author{Maximilian P. E. Lock}
\affiliation{Atominstitut, TU Wien, Stadionallee 2, 1020 Vienna, Austria}
\affiliation{Vienna Center for Quantum Science and Technology (VCQ), TU Wien, Vienna, Austria}
\affiliation{Institute for Quantum Optics and Quantum Information (IQOQI), Austrian Academy of Sciences, 1090 Vienna, Austria}

\author{Marcus Huber}
\affiliation{Atominstitut, TU Wien, Stadionallee 2, 1020 Vienna, Austria}
\affiliation{Vienna Center for Quantum Science and Technology (VCQ), TU Wien, Vienna, Austria}
\affiliation{Institute for Quantum Optics and Quantum Information (IQOQI), Austrian Academy of Sciences, 1090 Vienna, Austria}

\author{Iva B\v rezinov\'a}
\email{iva.brezinova@tuwien.ac.at}
\affiliation{Institute for Theoretical Physics, TU Wien, Wiedner Hauptstra\ss e 8-10/136, 1040 Vienna, Austria}
\affiliation{Vienna Center for Quantum Science and Technology (VCQ), TU Wien, Vienna, Austria}

\date{\today}

\begin{abstract}

In recent years, new theoretical insights into decoherence and quantum measurements have emerged through the study of many-body dynamics in isolated quantum systems. It is now understood that the parameters and energy scales in system-environment interactions decisively affect how readily information spreads from a quantum system into its surroundings during a decoherence event. A popular choice for studying these effects is the framework of quantum Darwinism (QD), but so far few works have applied this to realistic many-body models. Inspired by experimentally-accessible setups, in this work we introduce a simple, flexible, numerically exact many-body model of a system broadcasting information into an environment: a 1D lattice of sites with hopping particles. We show that different choices of parameters lead to the recovery of known scenarios featuring different decoherence and QD effects, such as equilibration, revivals of coherence, and redundancy. In constructing this model we resolve the crucial issue of indistinguishability: we explain how to calculate the entropy of a fraction of the environment when said environment is composed of indistinguishable fermions or bosons (or lattice sites containing them). We then show that particle statistics can make a notable difference to the QD properties of the setup, with fermionic environments sometimes achieving redundancy much more readily than bosonic or site-based ones. Our work opens the door to much closer alignment between theoretical models and experimental tests of the dynamics of quantum measurements and the quantum-to-classical transition.

\end{abstract}
\maketitle

\section{Introduction}
\label{sec:introduction}

Since von Neumann's original model of measurement \cite{vonNeumann1930}, it has long been understood that measurement in its conventional, textbook form is an interaction between a quantum system and a classical one, often phrased as the two systems lying on either side of the so-called  \textit{Heisenberg cut} \cite{09WisemanMilburn}. This interaction is said to extract information from the quantum system and make it appear classical. But the precise location of the cut in the measurement process, the exact mechanism of the information transfer, and the true nature of the quantum-to-classical transition are all difficult to quantify. Reasons for this difficulty include conceptual problems in quantum foundations \cite{96BuschLahtiMittelstaedt, bell_against_measurement}, questions of which observers count as classical \cite{18FrauchigerRenner, baumann20, relano20, 25RivlinEngineerBaumann}, conflicts with the laws of thermodynamics \cite{szilard1929, 64Szilard, peres80, 83Zurek, 13Hormoz, QThermoReview16, 20GuryanovaFriisHuber, carroll2021energy, 23MohammadyMiyadera}, and ambiguities in the thermodynamic costs of acquiring information from measurements \cite{17ElouardHerreraAuffeves, 23TarantoBakhshinezhadBluhm, debarba2024broadcasting, 25MohammadyBuscemi, candeloro26, ballesteros26}.

Part of the issue is that there is no consensus on how to describe measurements dynamically. In recent years, many models have been proposed to study this \cite{allahverdyan13, 23ArtiniPaternostro, 25ArtiniLoMonacoPaternostro, 25WallsBlossFord, 25LatuneElouard, schwarzhans25, schwarzhans26, engineer26}. Each of them invoke decoherence theory in some form, making it the dominant paradigm to study this and many related problems \cite{zurek_1981_pointer, zurek_1982_environment, joos_emergence_1985, schlosshauer_2005_decoherence, 22Zurek}. Decoherence describes a quantum-to-classical transition as an interaction between a quantum system and a surrounding environment, that destroys `quantumness' in the system and leaves only the classical information in a particular basis called the pointer basis. The emergence of classicality is then often associated with the emergence of a mixed state $\rho$ from the original pure state $\ket{\psi}$ \cite{luders_uber_1950, joos_emergence_1985}. Measurement can then be thought of as a decoherence event where the environment includes some form of device, apparatus, or observer.

As decoherence theory developed, further requirements for the emergence of classicality were formulated via a framework called quantum Darwinism (QD) \cite{09Zurek}. Within QD, it is required that not only should the environment be conditioned on the pointer states of the systems, but also the information about the pointer states should be \textit{persistently} stored in the environment, and the storage should be \textit{redundant}, meaning many different fractions of the environment should encode the same information. Key to this idea is the notion of \textit{objectivity}: every classical observer of the system should observe the same measurement outcome \cite{15HorodeckiKorbiczHorodecki}. The presence of these features of QD has been cited as one of the key indicators of the emergence of classicality in a decoherence event \cite{21Korbicz}.

QD has been the subject of intense study, with many recent theoretical works dedicated to its foundations and implications for other fields \cite{ollivier_objective_2004, blume05, riedel12, galve2016non, 17MironowiczKorbiczHorodecki, 19LeOlayaCastro, 20LeOlayaCastro, 22TouilYanGirolami, duruisseau23, chisholm23, doucet24, chisholm24, chisholm_emergence_2026, kiely26} and also a few experimental tests \cite{18CiampiniPinnaPaternostro, 25ZhuSaliceTouil}. Alongside these works sits a growing literature exploring quantum Darwinism in many-body models \cite{15GiorgiGalveZambrini, 16ZwolakRiedelZurek, 21MirkinWisniacki, 21RyanPaternostroCampbell, 21KwiatkowskiCywinskiKorbicz}, which study the dynamics that lead to the persistency, redundancy, and objectivity criteria of QD. But so far, these works have been limited to idealised many-body models, even numerically. And in particular, no previous work on this subject has studied the effects of indistinguishability on QD effects in many-body models, which severely hinders the ability of said models to connect to achievable experimental setups. 

In this work, we close this gap by outlining precisely how to consider the effects of indistinguishability in the study of QD in closed many-body systems. This is done in the context of a new closed-system, many-body model we propose that is numerically exact, experimentally-inspired \cite{bloch05, bloch_quantum_2012, gross_quantum_2017, 26KendrickKaleGreiner}, flexible, and highly tunable: a 1D lattice consisting of sites that can be occupied by interacting fermions or (hard-core) bosons that hop between the sites.

The key metric used in the study of QD is the quantum mutual information (QMI) ~\cite{blume05, 09Zurek}, which relies on a calculation of the entropy of a \textit{fraction} of the environment. The main issue that indistinguishability raises for this is that it becomes unclear what such a fraction means precisely in the context of a QMI calculation. Here, we show how to calculate this when the fraction being considered is composed of fermions or hard-core bosons, or different sites in the lattice. In particular, we demonstrate clear differences between the QMI calculated based on `particle entanglement' (i.e.\ traces over particles) as compared to `orbital entanglement' (i.e.\ traces over single-particle orbitals, here represented by sites in the lattice). We find that the behaviour of the former depends on the particle statistics, while the latter does not.

We also find that for certain Hamiltonian parameter choices, the exchange symmetry of the particles can affect QD effects such as redundant encoding. In agreement with previous research on simpler models, we show that we can tune the parameters of our model to produce dynamics that leads to equilibration, the revival of coherences, long-lasting redundancy, and other well-known properties of decoherence, QD, and the quantum measurement process. 

In particular, previous works have shown that intra-environmental interactions are detrimental to the goal of persistently storing information in the environment \cite{riedel12,21RyanPaternostroCampbell,doucet24}, especially when the goal is for the information to be \textit{redundantly encoded} in the environment. If the strength of the environment-environment interactions is similar to those of the system-environment interactions, it should not be possible to achieve redundancy in the QD sense. This can also be thought of as a statement about the timescales of redundant encoding versus scrambling, with for instance \cite{riedel12} showing that weaker intra-environment interactions allow for the long-lived, redundant storage of information on intermediate timescales (though the environment still eventually thermalises and loses all information about the system on longer timescales).

This also has implications for any model that attempts to unitarily reproduce the features of quantum measurements. In such a model, a quantum particle's microstate must effect a macroscopic change in its surrounding environment. As outlined above, for this to happen, we require that the system-environment interactions dominate the environment-environment interactions in the Hamiltonian, or at least that we consider timescales over which this is the case. (An alternative would be that we supply a specifically designed detector with an appropriate amplification mechanism, with all of the out-of-equilibrium thermodynamic resources that implies \cite{20GuryanovaFriisHuber, schwarzhans26, 25LatuneElouard, ballesteros26}, but this is not considered here.) Indeed, thermodynamic and statistical-mechanical aspects such as closed-system equilibration \cite{rigol08, 16GogolinEisert, Meier25} and the eigenstate thermalisation hypothesis \cite{dalessio2016, deutsch2018eigenstate} have not been considered to any great degree in studies of the properties of QD and their relation to measurement until quite recently \cite{schwarzhans25, engineer26, Cao26}.

On a separate note, in a recent series of papers \cite{chisholm23, chisholm24, chisholm_emergence_2026, kiely26} many properties defined in QD such as redundancy and objectivity were shown to have subtleties in their definitions -- they were shown to be ambiguously defined when different parts of the environment have different properties. This is important as redundancy guarantees that independent fractions of the environment come to the same conclusion about the state of the system, which is a necessary condition for objectivity, and the emergence of classicality is often associated with objectivity in this sense \cite{15HorodeckiKorbiczHorodecki}. This ambiguity can result in quantifiably different results for different averagings over parts of the environment. The work presented here on indistinguishability is a crucial development in this series of clarifications and disambiguations of QD concepts.

The numerical model we present here is inspired by the fact that one of the most powerful platforms currently available to study quantum many-body systems experimentally are ultracold atom quantum simulators \cite{bloch05,bloch_quantum_2012, gross_quantum_2017, 26KendrickKaleGreiner}. These platforms feature high tunability of both the strength of the external quenches and the interactions between the particles, and also facilitate advanced, precise techniques with which the quantum many-body states can be probed (see e.g.~\cite{kuhr_quantum-gas_2016}). Ultracold atom simulators typically consist of indistinguishable bosonic or fermionic atoms, while well-defined subsystems can be realised through `impurity' atoms of different species. This underlines the importance of considering indistinguishability when designing experimentally-testable QD models. With this study of the impact of indistinguishability on the emergence of quantum Darwinistic classicality for realistic many-body dynamics, our work opens the powerful platforms of quantum simulators to studies of fundamental questions concerning the quantum measurement process.

This paper is structured as follows: In Sec.~\ref{sec:setup} we will introduce the model system and analyse its basic properties. In Sec.~\ref{sec:qmi} we will discuss how to obtain the QMI between the impurity and the environment for the different partial traces. We will present our results in Sec.~\ref{sec:results} and conclude in Sec.~\ref{sec:conclusions}. The appendix contains further results underpinning those presented in the main text.

\section{The Setup}
\label{sec:setup}

\subsection{The system and its Hamiltonian}
\label{subsec:hamil}
\begin{figure}[t]
    \centering
    \scalebox{0.4}{\begin{tikzpicture}[
    scale=1.15,
    >=Latex,
    line cap=round,
    line join=round
]

\definecolor{impuritypink}{RGB}{149,13,69}
\definecolor{bathblue}{RGB}{57,60,129}
\definecolor{interactiongreen}{RGB}{83,97,66}
\definecolor{hoppingorange}{RGB}{164,85,0}

\draw[
    gray!75,
    line width=1pt,
    smooth,
    samples=500,
    domain=0:{16*pi/2.7}
]
plot (\x,{-1.6*(sin(deg(2.7*\x)))^2+0.6});

\coordinate (impL) at ({(pi/2 + 0*pi)/2.7}, -0.3);
\coordinate (bath1) at ({(pi/2 + 1*pi)/2.7}, -0.3);
\coordinate (bath2) at ({(pi/2 + 4*pi)/2.7}, -0.3);
\coordinate (bath3) at ({(pi/2 + 7*pi)/2.7}, -0.3);
\coordinate (bath4) at ({(pi/2 + 10*pi)/2.7}, -0.3);
\coordinate (bath5) at ({(pi/2 + 11*pi)/2.7}, -0.3);
\coordinate (impR) at ({(pi/2 + 15*pi)/2.7}, -0.3);
\coordinate (impL2) at ({(pi/2 + 0*pi)/2.7}, -0.6);
\coordinate (impR2) at ({(pi/2 + 15*pi)/2.7}, -0.6);


\shade[ball color=impuritypink]
(impL) circle (0.24);

\shade[ball color=bathblue] (bath1) circle (0.2);
\shade[ball color=bathblue] (bath2) circle (0.2);
\shade[ball color=bathblue] (bath3) circle (0.2);
\shade[ball color=bathblue] (bath4) circle (0.2);
\shade[ball color=bathblue] (bath5) circle (0.2);

\shade[ball color=impuritypink]
(impR) circle (0.24);


\draw[
    <->,
    line width=1.6pt,
    color=interactiongreen,
    opacity=1,
    >=Stealth
]
($(impL)+(0.23,-0.02)$)
--
($(bath1)+(-0.23,-0.02)$);

\node[
    align=center,
    text=interactiongreen,
    font=\LARGE
]
at ({(pi/2 + 0.5*pi)/2.7 +0.1 },1.7)
{
impurity-\\
environment\\
interaction
};


\node[
    align=center,
    text=bathblue,
    font=\LARGE
]
at ({(pi/2 + 4*pi)/2.7},1.6)
{
environmental\\
particles
};

\pgfmathsetmacro{\xminA}{(pi/2 + 7*pi)/2.7}
\pgfmathsetmacro{\xminB}{(pi/2 + 8*pi)/2.7}

\draw[
    ->,
    line width=1.3pt,
    color=hoppingorange,
    color=hoppingorange,
    opacity=1,
    >=Stealth
]
({\xminA+0.07},0.2)
to[
    out=75,
    in=110,
    looseness=2.8
]
({\xminB},-0);

\pgfmathsetmacro{\xminA}{(pi/2 + 6*pi)/2.7}
\pgfmathsetmacro{\xminB}{(pi/2 + 7*pi)/2.7}

\draw[
    ->,
    line width=1.3pt,
    color=hoppingorange,
    opacity=1,
    >=Stealth
]
({\xminB-0.07},0.2)
to[
    out=110,
    in=75,
    looseness=2.8
]
({\xminA},0);

\node[
    text=hoppingorange,
    font=\LARGE
]
at ({(pi/2 + 7*pi)/2.7},1.6)
{hopping};

\draw[
    <->,
    line width=1.6pt,
    color=hoppingorange,
    opacity=1,
    >=Stealth
]
($(bath4)+(0.26,-0.02)$)
--
($(bath5)+(-0.25,-0.02)$);

\node[
    text=hoppingorange,
    font=\LARGE,
    align=center
] at ({(pi/2 + 10.5*pi)/2.7},1.7)
{intra-environmental \\
interaction};

\node[
    text=impuritypink,
    font=\LARGE
]
at ({(pi/2 + 15*pi)/2.7-0.1},1.23)
{impurity};


\node[
    text=impuritypink,
    font=\LARGE
] (superposition)
at (9.5,-1.9)
{superposition};

\draw[
    ->,
    >=Stealth,
    line width=1.6pt,
    color=impuritypink
]
($(superposition.west)+(-0.05,0)$)
to[out=180,in=-40]
($(impL)+(0.25,-0.25)$);

\draw[
    ->,
    >=Stealth,
    line width=1.6pt,
    color=impuritypink
]
($(superposition.east)+(0.05,0)$)
to[out=0,in=-140]
($(impR)+(-0.25,-0.25)$);

\end{tikzpicture}}
    \caption{A sketch of the model consisting of a discrete 1D lattice with hard-wall boundaries and $M_{\mathrm{S}}$ sites (the sites are suggestively pictured as potential wells but in our model they are simply treated as abstract sites). The system contains $N_\mathrm{E} = M_{\mathrm{S}} / 2$ indistinguishable particles (fermions or hard-core bosons), which can hop from site to site and interact with each other via nearest-neighbour interactions. We refer to these particles as the \textit{environment}. Additionally, the system contains one distinguishable particle, which we call the \textit{impurity}, that is initially in a superposition of being at the left- and right-most sites. It interacts with the environmental particles in the ensuing dynamics, but it can only be at site 1, site $M_{\mathrm{S}}$, or a superposition of the two.}
    \label{fig:fig1}
\end{figure}
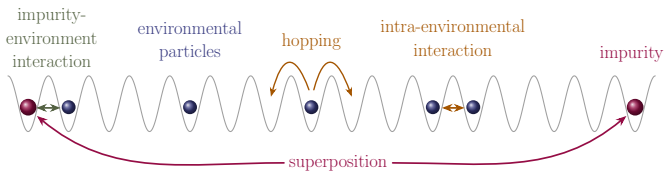

Our system is inspired by the ultracold atom experiments with optical lattices \cite{bloch05,bloch_quantum_2012,gross_quantum_2017, 26KendrickKaleGreiner}, which has seen much progress in recent years. Specifically, we consider a one-dimensional (1D) lattice where particles move via nearest-neighbour tunnelling from site to site -- see Fig.~\ref{fig:fig1}. Schematically, this lattice can be thought of as a chain of sites, each of which can be either unoccupied, or occupied by a single particle, with the particles able to hop between the sites. (This is similar to a Fermi-Hubbard model \cite{10Esslinger, 26KendrickKaleGreiner}.) The distribution of particles in the sites is the key observable we will study in this work, along with other quantities derived from it.
We denote the number of sites as $M_\text{S}$ and assume hard wall boundary conditions at the boundary of the lattice. In this lattice, $N_\text{E} = M_\text{S}/2$ indistinguishable particles can move around by hopping to adjacent sites. These particles are collectively called the environment, $E$. We consider these particles to be either (spin-polarised) fermions or hard-core bosons. Hard-core in this context means that, despite obeying bosonic exchange symmetry, the particles repel each other in a similar manner to fermions, such that only one particle can occupy a given site at once.

One additional particle is present, which we call the impurity, $I$. Its own exchange symmetry is irrelevant outside of being distinguishable in some way from the environment particles (e.g.\ one can think of it as a different species of fermion). It is constrained to only be present at either site 1, site $M_\text{S}$, or a superposition of the two, and is always initially in an equal superposition. Interactions between $I$ and $E$ will change how the environment particles are distributed around the sites in the lattice. In a more realistic scenario, one could assume that the impurity is a heavy atom with much smaller hopping matrix elements $J_\text{I}$ than the hopping matrix elements of the environmental particles $J_\text{I}\ll J_\text{E}$. In this case, the dynamics presented in Sec.~\ref{sec:results} would be valid on time scales $\tau \ll \hbar/J_\text{I}$ and the system would ultimately thermalise on longer time scales \cite{22KourehpazDonsaBrezinova}.

Here, however, we consider a simplified set of scenarios governed by a small number of Hamiltonian parameters with no impurity hopping. First, the absolute energy scale of the problem is determined by the parameter $J_\text{E}$ in the environmental self-Hamiltonian. This parameter is not varied -- it is used as the unit of energy and thus set to 1 along with $\hbar$ throughout this work. (And the impurity self-energy $J_\text{I}$ is set to 0 throughout.)

Next, interactions between environment particles on different sites are determined by the parameter $W_\text{EE}$, which governs the probability of hopping between sites. We only allow particles to hop between adjacent sites. In both the fermionic and hard-core bosonic cases, on-site interactions are prohibited, so there are no on-site intra-environment interactions in our Hamiltonian.

We then consider different strengths of the interactions between the impurity and the environment particles by changing the parameter $W_\text{IE}$. This parameter controls how much the impurity attracts the environment particles towards it. The range over which this attraction is present is given by $i_\text{max}$, and the extent to which distance from the impurity attenuates the interaction is governed by the set $\{s_i\}$.

The total Hamiltonian is thus expressed in second quantisation as 
\begin{equation}\label{eq:H_total}
    \hat H = \hat H_\text{E} + \hat H_\text{IE},
\end{equation}
with
\begin{align}\label{eq:H_E}
        \hat{H}_\mathrm{E} = & - J_\mathrm{E} \sum_{j=1}^{M_\mathrm{S} -1} \big (\hat{c}_{j+1}^\dagger \hat{c}_{j} \,+ \, \hat{c}_{j}^\dagger \hat{c}_{j+1} \big ) \\ 
        & +\, W_{\mathrm{EE}} \, \sum_{j=1}^{M_\mathrm{S}-1}
        \hat{N}_{j+1}\hat{N}_{j},
\end{align}
\begin{equation}
    \hat{H}_{\mathrm{IE}} = W_{\mathrm{IE}}  \sum_{i=0}^{i_{\mathrm{max}}} s_i \big( \sum_{j=1}^{M_\mathrm{S}-i} \hat{n}_j \hat{N}_{j+i} + \hat{n}_{j+i} \hat{N}_j \big ).
     \label{eq:H_IE}
\end{equation}
In the above equations, for each site $j$, $\hat c_j$ ($\hat c_j^{\dagger}$) is the annihilation (creation) operator of an environmental particle, $\hat N_j = \hat c_j^\dagger c_j$ is the number operator of environmental particles, and $\hat n_j$ is the number operator of the impurity. In Eq.~\ref{eq:H_IE} we assume that the impurity interacts with the environment particles over several sites as parametrised by the parameters $\{s_i\}_{i=1}^{i_\text{max}}$. For a given value of $i_\text{max}$, each $s_i$ is kept constant. They are chosen such that the interaction decreases almost linearly with distance from the site of the impurity. (In a laboratory setting, interactions reaching over several sites can be realised through dipolar atoms \cite{chomaz_dipolar_2023}.) See Tab.~\ref{tab:par_defs} for a summary of all of the parameters used in this work, and Tab.~\ref{tab:parameterdetails} in the Appendix for a more detailed breakdown of the parameters.

That leaves $W_\text{EE}$, $W_\text{IE}$, and $i_\text{max}$ as the three parameters that we vary in this work. For each of them, two possible values are considered (and each value of $i_\text{max}$ has a single associated set of $s_i$ values), resulting in eight sets of parameters (see Tab.~\ref{tab:parametersets} below). Then for each of the eight Hamiltonians (each defined by a given set of parameters) we study the dynamics for fermions and for hard-core bosons, resulting in 16 sets of results overall.

\begin{table}[t]\label{tab:par_defs}
\centering
\begin{tabular}{l l}
\toprule
\textbf{\(M_\mathrm{S}\)} & Number of lattice sites \\
\textbf{\(N_\mathrm{E}\)} & Number of bath particles in the system \\
\textbf{\(J_\mathrm{E}\)} & Environment particle self-energy (set to 1 throughout) \\
\textbf{\(W_{\mathrm{EE}}\)} & Next-neighbour interaction between env. particles \\
\textbf{\(W_{\mathrm{IE}}\)} & On-site interaction between impurity and bath \\
\textbf{\(s_i\)} & Attenuation parameters for the long-range \\
                    & impurity–environment interaction \\
$i_{\mathrm{max}}$ & Maximum distance over which the impurity interacts\\
& with the environment particles.  $i_{\mathrm{max}}\in[0,\frac{M_\mathrm{S}}{2}-1]$\\
\bottomrule
\end{tabular}
\caption{Overview of Hamiltonian parameters. Of these, $W_\text{EE}$, $W_\text{IE}$, and $i_\text{max}$ are the three that we vary to create the 8 cases we consider for each particle type. $J_I=0$ throughout. }
\end{table}

The initial state of the entire lattice is $|\Psi(0)\rangle$, and it evolves under the Hamiltonian $\hat{H}$ in Eq.~\ref{eq:H_total}. We propagate the state with the time-dependent Schrödinger equation $i\partial_t|\Psi(t)\rangle=\hat{H}|\Psi(t)\rangle$ through exact diagonalisation.

We first solve the time-independent Schrödinger equation. As a basis for diagonalisation we use the configuration space of second quantisation with many-body basis states given by $\{|n_1\dots n_{M_\text{S}}\rangle\otimes| N_1\dots N_{M_\text{S}}\rangle\}$, where $N_j=\{0,1\}$ denotes whether site $j$ is occupied by an environmental particle, and $\sum_{j=1}^{M_\text{S}}N_j = N_\text{E}$. Since there is only one impurity, we have $n_k=1$ for one $k$ in each configuration and all other $n_{j\neq k}=0$. Conveniently, we also have that the Hamiltonian is block-diagonal with respect to the impurity position since the impurity does not move. We represent the Hamiltonian in this basis and diagonalise it to obtain the eigenvalues $\{\mathcal{E}_n\}_{n=1}^{\dim(\mathcal{H})}$ and the many-body eigenstates $\{|\mathcal{E}_n\rangle\}_{n=1}^{\dim(\mathcal{H})}$.

Regarding the dimension of the Hilbert space, we have a total dimension of $\text{dim}(\mathcal{H}) = M_\text{S}\times \text{dim}(\mathcal{H}_\text{E}) = M_\text{S}\times \binom{M_\text{S}}{N_\text{E}}$, which for the parameters we consider in the results resolves to 205920. But since we only consider cases where the impurity does not move, the total Hilbert space dimension we work with is effectively only  $2\times \binom{M_\text{S}}{N_\text{E}}$, which resolves to 25740.

Note that, since the configuration space is exactly identical for both the fermions and the hard-core bosons, both the spectrum and the eigenstates in the configuration space basis are exactly equal for both species. This is crucial if we want to pin down the role of the statistics of the particles on the measurement dynamics. The only difference here between fermions and bosons is that the basis states $\{| N_1\dots N_{M_\text{S}}\rangle\}$ represent either Slater determinants for the fermions or matrix permanents for the bosons \cite{18NegeleOrland}. Hence, all observables that are not sensitive to particle exchange, such as the spatial particle density distribution, will be exactly equal for both species. However, importantly, other observables that are sensitive to particle exchange will differ. The quantum mutual information (see Sec.~\ref{sec:qmi}), central to our study, is one such observable.

The objective in this study is to observe the dynamics of the environment, in particular how this decoheres the impurity and essentially `measures' its position. While the structure of the Hamiltonian guarantees the emergence of a well-defined decoherence pointer basis, the presence of interactions between bath particles prevents the formation of the ``singly-branching'' state structure necessary for zero discord, and thus perfect Darwinism~\cite{blume05,blume2006quantum,touil2024branching}. Indeed, the form of two-body Hamiltonians necessary for this condition are well-characterised in~\cite{duruisseau23,doucet24}. However, since some amount of intraenvironment interaction is to be expected in any real system, we relax this requirement and investigate the extent to which the environment may nonetheless redundantly encode the position of the impurity.

\subsection{The initial state and its time evolution}\label{sub:sec_init_psi}

We choose as the initial state $|\Psi(0)\rangle$ a state where the impurity is in an equal spatial superposition between the left-most site $1$ and the right-most site $M_\text{S}$, and the environmental particles are in the ground state of $\hat H_\text{E}$, denoted by $|\phi_1\rangle_\text{E}$.

To simplify the notation in the following, we will use $|\text{L}\rangle_\mathrm{I} = |10\dots0\rangle$ for the impurity state localised at the left-most site and $|\text{R}\rangle_\mathrm{I} = |0\dots01\rangle$ for the impurity state localised at the right-most site. The initial state of the impurity is then given by
\begin{equation}\label{eq:psi_I_0}
    \ket{\psi(t=0)}_\mathrm{I} = \frac{1}{\sqrt{2}} \big ( \ket{\text{L}}_\mathrm{I} + \ket{\text{R}}_\mathrm{I} \big ),
\end{equation}
and the total state is
\begin{equation}\label{eq:Psi_tot_0}
    \ket{\Psi(0)} = |\psi(0)\rangle_\text{I} \otimes \ket{\phi_1}_\text{E}.
\end{equation}
The time evolution of this state is due to the interaction $\hat H_\text{IE}$ between the impurity and the environmental particles. Upon spectral decomposition into the eigenstates of the total system, the state is evolved in time according to 
\begin{equation}\label{eq:Psi_tot_t}
    \ket{\Psi(t)} = \sum_{n=1}^{\text{dim}(\mathcal{H})} C_n e^{-i\mathcal{E}_nt}|\mathcal{E}_n\rangle.
\end{equation}

Since the impurity is being kept fixed at its position(s) (i.e.~$\hat H_\text{I}=0$) the time-evolved state can always be decomposed in the following way:
\begin{equation}\label{eq:Psi_tot_cond}
    \ket{\Psi(t)} = \frac{1}{\sqrt{2}}\ket{\text{L}}_\text{I}\ket{\phi_\text{L}(t)}_\text{E}+\frac{1}{\sqrt{2}}\ket{\text{R}}_\text{I}\ket{\phi_\text{R}(t)}_\text{E}.
\end{equation}
Then note that at $t=0$, $\ket{\phi_\text{L}(0)}_\text{E} = \ket{\phi_\text{R}(0)}_\text{E} = \ket{\phi_1}_\text{E}$.

\subsection{Some useful metrics}\label{sub:sec_useful_metrics}

The objective of the study will be to determine how effectively the dynamics cause the environmental particles to encode the information about the position of the impurity. This can be thought of as the environment `measuring' the impurity in the position basis, or the environment inducing decoherence on the impurity with position as the pointer basis. We now introduce some useful metrics to characterise this.

First, we introduce the so-called effective dimension, $d_\text{eff} = 1/\sum_n|C_n|^4$ (see e.g.~\cite{11Short}). This quantity encapsulates the effective number of states $|\mathcal{E}_n\rangle$ that contribute to the evolution of $|\Psi(0)\rangle$. Also known as the inverse participation ratio, it can also be thought of as the effective number of dimensions of Hilbert space that the dynamics occurs in, which can in general be much smaller than the total dimension of the Hilbert space.

Next, we introduce a measure of the extent to which decoherence has occurred. We define 
\begin{equation}\label{eq:D(t)}
    D(t) = |_\text{E}\langle \phi_\text{L}(t)|\phi_\text{R}(t)\rangle_\text{E}|^2,
\end{equation}
as a decoherence parameter (see e.g.~\cite{relano20}). In a perfectly decohered scenario, it is possible to perfectly predict the state of the system by observing the state of the environment, so the environment state commensurate with the particle being on the left and environment state commensurate with the particle being on the right should have perfect distinguishability -- their overlap should be precisely zero. Note that we call $D(t)$ a \textit{de}coherence parameter, despite the fact that it being \textit{lower} is a sign of more decoherence in the lattice.

Next we introduce the reduced density matrix of the impurity obtained by tracing out all environmental particles 
\begin{equation}\label{eq:rho_I}
    \hat \rho_\text{I}(t) = \text{Tr}_{1\dots N_\text{E}}|\Psi(t)\rangle \langle \Psi(t)|.
\end{equation}
In a scenario with perfect decoherence, this quantity is diagonal in the `site basis' -- the basis of single-particle states $|j\rangle$ localised at individual sites. In such a case, we have $\langle i|\hat \rho_\text{I}(t)|j\rangle = 1/2 \delta_{ij}$ for $i=1,M_\text{S}$ and zero otherwise. When decoherence is not perfect, off-diagonal elements will be present.
Note also that the number operators of the impurity on individual sites, $\hat n_j$, commute with the total Hamiltonian: $[\hat n_j,\hat H]=0\,\,\forall j$. This criterion is necessary criterion for the pointer basis to be \textit{stable} -- to not vary over time. Throughout this work we do not consider scenarios where the environment stably records information about the impurity even after that information is no longer present in the impurity -- we only consider \textit{non-invasive} measurements \cite{20GuryanovaFriisHuber}.

Within quantum Darwinism, the broadcasting of information into the environment is usually quantified with a metric called the quantum mutual information (QMI) \cite{nielsen_chuang_2010}. Given a subsystem $A$ and a subsystem $B$, the mutual information between $A$ and $B$ is given by 
\begin{equation}
    I[A:B] = S(\hat\rho_\text{A})+S(\hat \rho_\text{B}) - S(\hat \rho_{AB})
    \label{eq:QMI1}
\end{equation}
with the von Neumann entropy $S=-\text{Tr}[\hat \rho \log_2\hat \rho]$. (We use base-2 for the logarithm as our system of interest, the impurity, can only be in two states, left or right, and hence can be thought of as a qubit.)

$I[A:B]$ is never negative, with $I[A:B]=0$ if and only if $A$ and $B$ are uncorrelated, i.e.~$\hat \rho_\text{AB}=\hat \rho_A\otimes\hat \rho_B$. The entropy of the reduced density matrix of the impurity, Eq.~\ref{eq:rho_I}, is $S[\hat \rho_I(t)]$. This is upper-bounded by 1, with $S[\hat \rho_I(t)]=1$ for $D(t)=0$. The mutual information between the impurity and the entire environment is $I[\text{I}:\text{E}] = S(\hat \rho_\text{I}) + S(\hat \rho_\text{E})= 2S(\hat \rho_\text{I})$, and its upper bound is 2. This quantity is the key metric we study in most of the Results section Sec.~\ref{sec:results}.

\section{Quantum mutual information for indistinguishable particles}\label{sec:qmi}

The key prediction of the quantum Darwinism framework is that information is stored in the environment in a \textit{redundant} way, meaning that different parts of the environment contain the same information about the system being measured or decohered. This notion is usually quantified by dividing the environment $\text{E}$ into fractions $\text{F}$, and then assessing how much information about the system is contained in each fraction.  The metric used for this is the QMI $I[\text{I}:\text{F}]$ of Eq.~\ref{eq:QMI1}. The QMI is usually plotted as a function of the \textit{size} of the fraction to discern how information is distributed in the environment. For instance, a quantity called the redundancy, $R$, can be defined as the inverse of the smallest fraction size for which the QMI is approximately 1 (with 1 being the value of $S(\hat \rho_I)$).

In our case, this metric is problematic. Although the impurity remains distinguishable from the environment, the environment particles are now indistinguishable from each other. Hence, in contrast to the usual spin systems analysed in studies of QD (see e.g.~\cite{ollivier_objective_2004, blume05, riedel12, duruisseau23}), one cannot uniquely define a fraction of the environment in the conventional QD sense \cite{ding_concept_2021}. One can only consider the QMI for an averaging over all possible fractions of a certain size, but even then there is no one unique way to perform this averaging. One of the main contributions of this work is that we consider two different ways to sensibly define fractions of the environment when the fractions are indistinguishable in some sense: averaging over particles, and averaging over orbitals (which here corresponds to sites). In this section, we describe how to obtain the QMI for both cases.

First we consider particle averaging. This involves tracing over environment particles to obtain $p$-particle reduced density matrices ($p$RDMs) \cite{coleman_structure_1963} for different numbers of particles $p\leq N_\text{E}$. The resulting $p$RDMs each pertain to a fixed number of particles, which are representative of all the choices one could have made in particle selection. This case is often associated with `particle entanglement' in the literature \cite{amosov_spectral_2017,galler_orbital_2021,ernst_mode_2024}. 

Site-averaging, by contrast, involves evaluating reduced density matrices for sites. It is more complicated than the particle-based $p$RDMs, since the particles can hop in and out of the corresponding fractions. The sites can thus be thought of as constituting an open quantum system, with the site-based reduced density matrices representing states in Fock space with no fixed particle number. Furthermore, because the sites are in general different from each other (e.g.~some are impacted by the $W_\text{IE}$ interaction and some are not), one must specify exactly which sites are being averaged over in any given calculation. Site-averaging is comparable to what is commonly referred to as `orbital entanglement' in the literature \cite{ding_concept_2021,galler_orbital_2021}.

Here we should contrast our work with a key set of recent results on environments in QD. In \cite{chisholm23, chisholm24}, it was shown that there are subtleties to considering fractions of the environment in QD when different parts of the environment have different properties, and it was explained how to properly account for averaging in such cases. This was all done in the distinguishable case, however, and so is not comparable to the averaging over indistinguishable particles we perform here, but that work does inform our calculations that involve averaging over sites.

\subsection{Particle-based averaging}\label{sec:particlebased}

To simplify notation, we introduce a combined creation operator $\hat C_J^\dagger = \hat c_{j_1}^\dagger\dots \hat c_{j_p}^\dagger$ with an ordered tuple of indices $j_1<\dots < j_p$ (and associated $\hat C_J$ annihilation operators). The unity in the $p$-particle Hilbert space is given by $\sum_J \hat C^\dagger_J\hat C_J$ with the sum over all ordered tuples $J$. We also introduce projection operators $\hat P_{JK} = \hat C_J^\dagger \ket{0}\bra{0} \hat C_{J\bigcup K}$, where $\ket{0}$ is the vacuum state.

Consider tracing over $p$ environment particles to produce the $p$RDM. This quantity can be defined as
\begin{eqnarray}\label{eq:D_Ip}
    \hat D_{\text{I},1\dots p} = \text{Tr}_{p+1\dots N_\text{E}}\left(|\Psi(t)\rangle\langle\Psi(t)|\right)\\
     = \sum_{JKL}\hat P_{JK}\ket{\Psi(t)}\bra{\Psi(t)}\hat P_{KL}^\dagger,
\end{eqnarray}
where $J=\{j_1<\dots<j_p\}$ is an ordered tuple of $p$ site indices, $L=\{l_1<\dots<l_p\}$ is a different ordered tuple, $J^C$ and $L^C$ are the complements of $J$ and $L$ respectively, and $K=J^C\bigcup L^C$. Note that with the help of the creation and annihilation operators, one can see that the reduced density matrices are sensitive to the statistics of the particles, i.e.~whether the creation and annihilation operators obey the commutation relations ($[\hat c_i,\hat c_j]=0$, $[\hat c_i^\dagger,\hat c_j^\dagger]=0$, $[\hat c_i,\hat c_j^\dagger]=\delta_{ij}$) or the anti-commutation relations ($\{\hat c_i,\hat c_j\}=0$, $\{\hat c_i^\dagger,\hat c_j^\dagger\}=0$, $\{\hat c_i,\hat c_j^\dagger\}=\delta_{ij}$).

Also note that the above definition $\hat D_{\text{I},1\dots p}$ is for a reduced density matrix that includes the impurity. The reduced density matrix for the $p$ environment particles without the impurity is given by $\hat D_{1\dots p} = \text{Tr}_\text{I}\hat D_{\text{I},1\dots p}$. For the particle-based QMI, we require both the entropy of $\hat D_{\text{I},1\dots p}$ and of $\hat D_{1\dots p}$. We obtain these entropies using diagonalisation. We will denote the QMI obtained using the particle-based partial trace by 
\begin{equation}\label{eq:pQMI}
    I_\text{part}[\text{I}:\hat D_{1\dots p}] = S(\hat \rho_\text{I})+S(\hat D_{1\dots p}) - S(\hat D_{\text{I},1\dots p}).
\end{equation}
The fraction of the environment which we are interested in studying in the QD context is then given by $F=p/N_\text{E}$. Note that the QMI shows the known (mirror) symmetry $I_\text{part}[\text{I}:\hat D_{1\dots p}] + I_\text{part}[\text{I}:\hat D_{1\dots N_{\text{E}}-p}] = 2$, and when the fraction contains exactly half the total number of environment particles, we have $I_\text{part}[\text{I}:\hat D_{1\dots N_{\text{E}/2}}]=1$. When distinguishing between the fermionic and bosonic cases we will refer specifically to $I_\text{part}^{\text{F}}$ and $I_\text{part}^{\text{B}}$.

\subsection{Site-based averaging}\label{sec:sitebased}

Next we consider the case of tracing over sites. This is more complicated because the sites are different from each other. Recalling that there are many different ordered tuples of sites of any given size, we must first specify which ordered tuple of sites $X=\{j_x\}_{x=1}^{m}$ we want to evaluate the density matrix of (where $|X|=m$ is the cardinality of the set $X$) and then we trace over its complement $X^C$: 
\begin{eqnarray}\label{eq:rho_S}
    \hat \rho_{\text{I},X} = \text{Tr}_{X^C}\left(|\Psi(t)\rangle\langle\Psi(t)|\right) \\
    =\sum_{JKL} \hat P_{JK}\ket{\Psi(t)}\bra{\Psi(t)}\hat P_{KL}^\dagger.
\end{eqnarray}
Here, $J$ is still an ordered tuple of site indices, but now it must be drawn from the set $X$ ($J\in X$), and is not constrained to have exactly $p$ indices. $L\in X$ and $K \in X^C$ have analogous definitions. 

$\hat \rho_{\text{I},X}$ is now a density matrix in Fock space mixing different numbers of particles in an incoherent way on $X$. Hence we can call it the site-based reduced density matrix. The incoherence is a consequence of particle number conservation, and leads to a block diagonal structure, such that $\hat \rho_\text{X}$ can be written as a sum over reduced density matrices of fixed particle number $\hat \rho_{\text{I},X}^p$:
\begin{align}
    \hat \rho_{\text{I},X} &= \sum_{p=0}^{|X|} \hat \rho_{\text{I},X}^p,
\end{align}
with the fixed-particle-number reduced density matrices defined as
\begin{eqnarray}
\hat \rho_{\text{I},X}^p = \sum_{J_p,K_p,L_p}
\hat P_{J_pK_p}\ket{\Psi(t)}\bra{\Psi(t)}\hat P_{K_pL_p}^\dagger
\end{eqnarray}
for $J_p$ an ordered tuple of site indices of length $p$ drawn from the set $X$ and similar for $L_p\in X$ and $K_p \in X^C$. 
As before, we then obtain the site-based reduced density matrix for the environment particles without the impurity by tracing $\hat \rho_{\text{I},X}$ over the impurity: $\hat \rho_{X}=  \text{Tr}_\text{I}\hat \rho_{\text{I},X}$.

The QMI is then given in the site basis by 
\begin{equation}\label{eq:sQMI}
    I_\text{site}[\text{I}:\hat \rho_X]= S(\hat \rho_\text{I}) + S(\hat \rho_{X})- S(\hat \rho_{\text{I},X}).
\end{equation}
At this point, a key difference between the site-based reduced density matrices $\hat \rho_{\text{I},X}^p$ and the $p$RDMs $\hat D_{\text{I},1\dots p}$ should be noted. Each matrix element of $\hat D_{\text{I},1\dots p}$ represents a configuration where the number of environment particles is $p$ and the number of sites they are distributed over is also $p$, but each matrix element of $\hat \rho_{\text{I},X}^p$ represents a configuration where $p$ environment particles can be distributed over any number of sites allowed in the set $X$. 

If $X$ is disjoint and one has to determine the matrix elements for configurations between these disjoint parts of $X$, the number of occupied sites over which one has to permute to connect these configurations is always the same. This means that, if the number of occupied sites is odd, the corresponding matrix element comes with an additional global $(-1)$ prefactor for fermions. In other words, the $\hat \rho_{\text{I},X}^p$ are equal for fermions and bosons up to a global $(-1)$ for the above described configurations. Since the $\hat \rho_{\text{I},X}^p$ for fermions and bosons can thus be transformed into one another through unitary transformations, they have the same spectrum. Hence, we can already describe one key result of this work: the QMI based on traces over sites does not depend on the particle statistics. It is only when tracing over particles that the exchange symmetry becomes important.

We will thus compare three different QMIs in the results section, the QMI based on particle traces $I_\text{part}^{\text{F}}$ for fermions, the QMI based on particle traces $I_\text{part}^{\text{B}}$ for bosons, and the QMI based on site traces $I_\text{site}$ (which is equal for both particle species).

Finally, in the case of traces over sites, there is an additional averaging we must perform. Sets of sites $X$ with the same cardinality $|X|$ can be different from each other (e.g.~a set of two sites near the impurity will have different properties to a set of sites near the middle of the chain). Hence, to understand the QMI for a given `size of the environment' in the QD sense, we must consider the QMI averaged over different sets $X$ of the same size. (Compare this to the averaging over environment fractions done in \cite{chisholm24}.)

The expression for this averaging is 
\begin{equation}\label{eq:sQMI_aver}
\langle I_\text{site}[\text{I}:\hat \rho_X]\rangle_{|X|} = \frac{1}{\binom{M_S}{|X|}} \sum_X I_\text{site}[\text{I}:\hat \rho_X] .  
\end{equation}
(Note that this expression is very expensive to calculate -- see Sec.~\ref{sec:results} for more details.) With this in mind, we can now say that the fraction of the environment we consider in the QD sense is the number of sites remaining after the trace, divided by the total number of sites: $F=|X|/M_{S}$. It is also important to note that in the case of traces over sites the QMI will not necessarily be mirror symmetric unless the fractions $F\geq 1/2$ are exactly chosen from the corresponding $X^C$.

In Sec.~\ref{subsec:red}, we will evaluate $I_\text{part}^{\text{F/B}}[\text{I}:\hat D_{1\dots p}]$ and $\langle I_\text{sites}(\text{I}:\hat \rho_X)\rangle_{|X|}$ for different choices of the Hamiltonian parameters.
\section{Results}
\label{sec:results}

Our objective is to study the decoherence dynamics of our impurity-environment lattice system to look for quantum Darwinism (QD) effects, and to understand the role indistinguishability plays in affecting these dynamics. To that end, we explore a number of different parameter regimes of the Hamiltonian for the initial state described earlier, in both the fermionic and bosonic cases, and we calculate the quantum mutual information (QMI) using both particle-based partial tracing (for both fermions and bosons) and site-based partial tracing.

In the following, we evaluate the dynamics of the total system for eight different sets of Hamiltonian parameters. The three parameters we vary are the nearest-neighbour interaction between the environmental particles ($W_\text{EE}$), the interaction strength between the impurity and the environment particles ($W_\text{IE}$), and the maximum range of the impurity-environment interaction ($i_\text{max}$), see Tab.~\ref{tab:parametersets} (and Tab.~\ref{tab:parameterdetails} in the Appendix for more details). Each of these three parameters has a low setting and a high setting for a total of eight parameter choices. We will discard two of the settings that lead to uninteresting dynamics as explained below, resulting in six scenarios to plot (or twelve overall when considering the fermionic and bosonic cases). 

Recalling that the energy scale is set by the environment self-interaction energy $J_E$ being equal to 1, we have that for $W_\text{EE}$, the low setting is 0 and the high setting is -1, for $W_\text{IE}$, the low setting is -1 and the high setting is -4, and for $i_\text{max}$, the low setting is 3 and the high setting is 7. This means that in all cases, we assume either attractive or vanishing interactions. Note that for each of the two values of $i_{\mathrm{max}}$ considered here, a certain set of $s_i$ values are used. In the short-range case, the three $s_i$ values are 0.8, 0.5, and 0.3, and in the long-range case, we have $s_i=1-i/8$. (See also the table in Appendix~\ref{app:params}.) This means that in the short-range case, the impurity impacts the environment over the three closest sites, and in the long-range case it impacts the environment over the seven closest sites.
\begin{table}[t]
\begin{center}
\begin{tabular}{c|c|c|c|c|l}
      Set & $W_{\mathrm{IE}}$ & $W_\text{EE}$ & $i_{\rm{max}}$ &$d_{\mathrm{eff}}$ & $\tau_{\mathrm{QSL}}$ \\
     \hline
     Persistent coherence (PC) & -1 &  0 & 3  &   3.62   &  10.47 \\
     Non-equilibrating (NE) & -1 & -1 & 3  &   8.34   & 5.06  \\
     Information scrambling (S)& -4 & -1 & 3  & 130.62   & 1.08 \\
     Redundantly encoding (R)& -4 &  0 & 3  &  22.90   & 1.28 \\
     PC, long range (PC-LR) & -1 &  0 & 7  &   4.19   & 5.59 \\
     NE-LR & -1 & -1 & 7   &  10.55   & 4.75 \\
     S-LR & -4 & -1 & 7   & 416.50   & 1.19 \\ 
     R-LR & -4 &  0 & 7   &  82.71   & 1.40 \\
\end{tabular}
\caption{Overview of the eight different parameter regimes considered in this work for fermions and bosons. Each of $W_{\mathrm{IE}}$, $W_{\mathrm{EE}}$, and $i_{\mathrm{max}}$ have a high and a low setting -- low and high $I$-$E$ interaction, low and high $E$-$E$ interaction, and short- and long-range $I$-$E$ interaction. Each of the eight parameter sets is assigned a name based on the dynamics they produce (see Sec.~\ref{sec:results}). Also shown here are the effective dimension $d_\text{eff}$ and the quantum speed limit time scale $\tau_\text{QSL}$ for each parameter set (for $M_\mathrm{S}=16$ and $N_\mathrm{E}=8$). Note that for each of the two values of $i_{\mathrm{max}}$ considered here, a fixed set of $s_i$ values are used -- see the table in Appendix~\ref{app:params}. Also note that PC and PC-LR results are not plotted due to them having insufficient decoherence dynamics.}
\label{tab:parametersets}
\end{center}
\end{table}
%
\subsection{Decoherence}\label{sec:decoherence}
In this subsection, we focus on the properties of the dynamics that are entirely independent of the statistics of the environmental particles. Focusing on the fermionic case in particular, we explore how the above parameter sets influence the dynamics of the decoherence parameter $D(t)$ from Eq.~\ref{eq:D(t)}. The key observable of the impurity-environment complex that we study is the fluctuations in density of the environment particles on the sites as a function of time. This observable, being experimentally accessible, gives many insights into the overall dynamics of the system.

Given the choice of our initial conditions, it should be expected that the environmental particles, initially congregated around the centre of the system, will be drawn towards the boundaries where the impurity is located. These dynamics are initiated by the attractive interaction with the impurity. Hence, if the impurity were located entirely on the left (right), the environment particles would be drawn towards the left (right). This means that the environment contains a record of the position of the impurity. When the impurity is in a superposition, there will be one state of the environment commensurate with the impurity being on the left, and one on the right. Hence it can be said that the density of the environment particles contains a record of the position of the impurity, acting as the measuring apparatus in the von Neumann model sense. 
For the first parameter set we study, consider the weaker impurity-environment interaction $W_\text{IE}=-1$, and vanishing environment-environment interaction $W_\text{EE}=0$, and short-range $i_\text{max}=3$ (the first line of Tab.~\ref{tab:parametersets}). For this, we can look at how $D(t)$ varies as a function of time. $D(t)$ reaches values close to zero for only very short times, after which there are long periods where $D(t)$ becomes larger again. This is caused by the particles being reflected from the hard-wall boundary and moving back towards the centre. It can also be thought of as the environment not having enough time to `explore' the impurity. Hence for this case, which we call the Persistent Coherence case (Set PC), we have $D(t)>0$ most of the time.

It is unsurprising that we have persistent coherence in this case because the effective dimension is very low. Despite the potential Hilbert space dimension of the $M_\text{S}=16$, $N_\text{E}=8$ setup being $\text{dim}(\mathcal{H})=25740$, the effective dimension of the dynamics is only $d_\text{eff} \approx 3$ (see Tab.~\ref{tab:parametersets}). This setup does not behave like a many-body system, and coherences can easily persist here -- we do not capture practically any decoherence behaviour. (Its long-range counterpart, Set PC-LR with $i_\text{max}=7$ has similar properties, too.) Both Set PC and Set PC-LR are not considered further in this work and are not plotted anywhere, leaving only six cases to consider for each of the two types of indistinguishable particles.

Interestingly, this situation changes as soon as $W_\text{EE}=-1$ -- when the environment-environment interactions is of the same magnitude as the energy scales $J_\text{E}$ and $W_\text{IE}$ (the second line of Tab.~\ref{tab:parametersets}). In this case, the environmental particles are dragged along with each other as soon as they start to move towards the impurity. The density of the environmental particles still moves back and forth between the boundaries of the system in a regular fashion, but the dynamics are slowed down by the interplay between interactions, the hopping and the prohibition on multiple occupation of sites. The overall effect is that we have an extended period of time where $D(t)$ falls and rises again, and we call the periods where $D(t)$ rises \textit{revivals}. Because of these revivals, we call this parameter set the Non-Equilibrating case (Set NE). (The long-range equivalent, Set NE-LR, has similar properties.) Though unlike with Set PC, this case still has at least some notable periods where $D(t)$ becomes close to zero. For completeness, we show in Appendix~\ref{app:sys_size}, Fig.~\ref{fig:dist_dim} that the time between the revivals increases with increasing system size.

\begin{figure}[t]
    \centering
    \hspace*{-0.6cm}

    \includegraphics[width=1\linewidth,trim=1cm 5.9cm 0.9cm 6.1cm, clip]{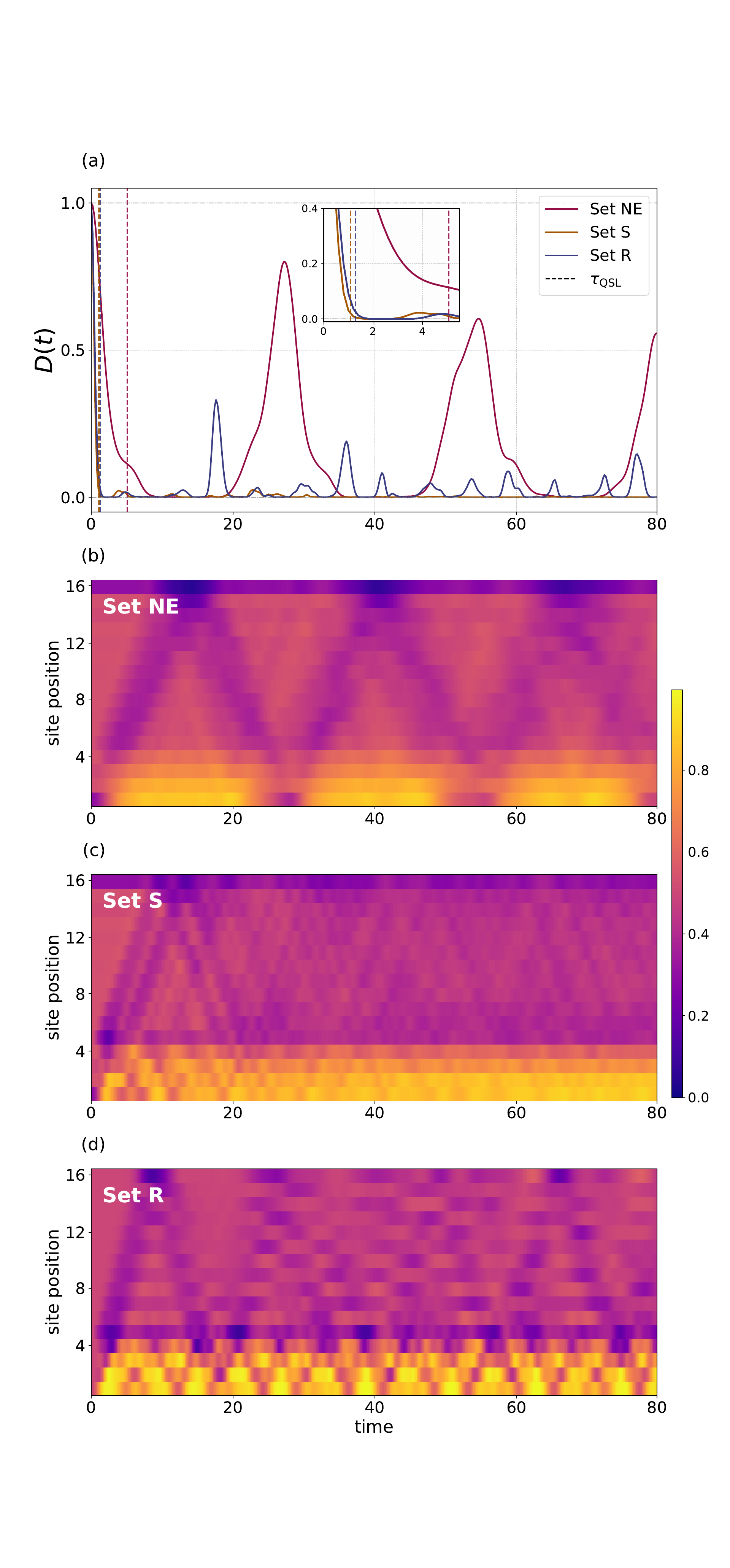}
    \caption{Comparison of three parameter sets for fermions with $M_\text{S}=16$ and $N_\text{E}=8$. See Tab.~\ref{tab:parametersets} for the parameter values. The vertical dashed lines in subfigure (a) mark $\tau_{\mathrm{QSL}}$ from Eq.~\ref{eq:tau_sql} for each case. (a) shows $D(t)$ as a function of time. (b-d) show, for each of the three cases, the probability of observing an environment particle at each lattice site over time, conditional on the impurity being at the left site, which are the diagonal elements of $\hat{D}^\text{L}_{1}$ defined below.}
    \label{fig:decoherence_density}
\end{figure}

In Fig.~\ref{fig:decoherence_density} (a), we show $D(t)$ as a function of time for fermions for the three short-range ($i_\text{max}=3$) parameter sets. The revivals are clearly visible in the NE line. We can clearly see that there is almost perfect decoherence $D(t)\approx 0$ for some periods of time, interspersed with periods of high revival where $D(t)>0$, with the strength of the revivals slowly decreasing with time. This is also reflected in the dynamics of the conditional density in Fig.~\ref{fig:decoherence_density} (b) (discussed in more detail shortly). Note also that the results in Fig.~\ref{fig:decoherence_density} (a) are exactly identical for fermions and for bosons, which will be important in the next subsection. 
We are also interested in the timescales over which these dynamics happen, and to that end we introduce the quantum speed limit $\tau_\text{QSL}$ (see e.g.~\cite{deffner17}):
\begin{equation}\label{eq:tau_sql}
\tau_{\mathrm{QSL}} = \text{max}\left\{ \frac{\pi \hbar}{ 2\Delta \hat H} , \frac{\pi \hbar }{2 \braket{\hat{H}}} \right\}
\end{equation}
with $\braket{\hat H}$ the energy expectation value of the total energy (relative to the ground state), and $\Delta \hat H$ its standard deviation. (In most of the cases we consider here it will turn out to be the $\braket{\hat{H}}$ term that dominates.) This time is called the quantum speed limit because it can be thought of as the fastest time scale possible for the quantum system in question to dephase \cite{deffner17}. This quantity is also plotted in Fig.~\ref{fig:decoherence_density} (a) as vertical lines for each parameter set, as it neatly characterises the time needed for the initial decoherence to happen (see also the inset). Hence it is clear from this that, for each parameter set, the observed decoherence dynamics is fast: $D(t)$ drops to almost zero within this fastest possible time scale.
Whenever $D(t)\approx 0$, we observe, as expected, that the density matrix of the impurity $\hat \rho_\text{I}$ is almost perfectly mixed, with two almost equal eigenvalues $\lambda_{1,2}\approx 1/2$ (see Fig.~\ref{fig:impurity_rdm} in Appendix~\ref{app:decor_dyn}). Again we can attribute the large revivals to the fact that for the NE case, the system is only weakly excited, having an effective dimension of only $d_\text{eff} \approx 8$ despite the aforementioned potential Hilbert space dimension of 25740. But unlike in the PC case, in the NE case the regular dynamics guarantee that $D(t)$ does remain very close to zero for extended periods of time. We even see a lack of fluctuations in the times when it is not experiencing a revival.

Next we consider the two cases where the magnitude of $W_\text{IE}$ is increased: $W_\text{IE}=-4$ with $W_\text{EE}=-1$ (the third row of Tab.~\ref{tab:parametersets}) and $W_\text{IE}=-4$ with $W_\text{EE}=0$ (the fourth row). For reasons that will become clearer in the next subsection, we respectively call these two sets Scrambling (Set S) and Redundant (Set R). In both cases, Fig.~\ref{fig:decoherence_density} (a) shows a rapid decay of $D(t)$ towards zero, and small amplitude fluctuations afterwards. And in both cases, we can speak of an equilibrated system, as deviations from the equilibrium value of $D(t) = 0$ are small and occur over short time scales. But the fluctuations are substantially smaller for Set S, due to its less regular spectrum, and much larger effective dimension. ($d_{\text{eff}}$ is known to control fluctuation size in closed-system equilibration \cite{Meier25}.) During periods of time where the environmental particles accumulate near the impurity, their overall dynamics reflects the dynamics of the decoherence parameter $D(t)$. That is, they have larger density fluctuations in the non-interacting case compared to the interacting case.  But broadly speaking, we can motivate the name `scrambling' by asserting that it is known that environment-environment interactions scramble information that the environment has about the system \cite{riedel12} (or in our case the impurity). 

Lastly in this section, in Fig.~\ref{fig:decoherence_density} (b-d) we also show the probability of observing an environment particle at each site as a function of time, \textit{conditional} on the impurity being on the left-most site (site 1), which we can think of as a conditional site density. This quantity is obtained from the one-particle reduced density matrix (1RDM) from Eq.~\ref{eq:D_Ip}: $\hat{D}_{\text{I},1}$. Because this object is expressed in a site basis, its diagonal elements correspond to the probability of observing an environment particle at each site. The \textit{conditional} 1RDM, $\hat{D}_{1}^{\text{L}}$ is then the top-left block of the $\hat{D}_{\text{I},1}$ matrix. The diagonal elements of $\hat{D}_{1}^{\text{L}}$ are plotted as a function of time in Fig.~\ref{fig:decoherence_density} (b-d) Note also that we re-normalise such that $\text{Tr}\left(\hat{D}_{1}^{\text{L}}\right)=N_\mathrm{E}$.

In each plot we do see much more probability concentrated on the left of the system at almost every timestep, reaffirming that the environment is tracking the position of the impurity. (The plots would look identical but mirrored for the right-conditional equivalent.) The decoherence dynamics are very visible in each of the plots, aligning well with $D(t)$ for each parameter set from Fig.~\ref{fig:decoherence_density} (a). For instance, we see clear revivals in Set NE, and almost no structure after decoherence in Set S. Overall these plots demonstrate that we are justified in treating this 1RDM as an indicator of how much information the environment contains about the position of the impurity.

Given these decoherence dynamics, we will now explore for the different cases (parameter sets NE, R, and S, Tab.~\ref{tab:parametersets}) how much information about the impurity is broadcasted into the environment and whether particle statistics plays a role. We will study their properties at a given time, and for this in each case we will hand-pick times where $D(t)\approx 0$.

\subsection{Redundancy in the QMI for environments with indistinguishable particles}\label{subsec:red}

Plotting the quantum mutual information as a function of an increasing fraction of the environment is the standard way to evaluate the QD properties of the post-decoherence system-environment state. The presence of a characteristic `plateau' in the middle of the plot would indicate that the predictions of QD are perfectly realised. As such, we now turn to analysing the effects that indistinguishability and different exchange symmetry have on the information stored in the environment as a function of fraction size. Below in Sec.~\ref{subsubsec:particle_entang}, we calculate the particle-based QMI using $I_\text{part}[\text{I}:\hat D_{1\dots p}]$ from Eq.~\ref{eq:pQMI} (which is associated with particle entanglement), and in Sec.~\ref{subsubsec:site_entang} we calculate the site-based QMI $I_\text{site}[\text{I}:\hat \rho_{X}]$ of Eq.~\ref{eq:sQMI} (associated with orbital entanglement).

\subsubsection{The mutual information with particle tracing}\label{subsubsec:particle_entang}

\begin{figure*}[t]
    \centering
    \includegraphics[width=1\textwidth]{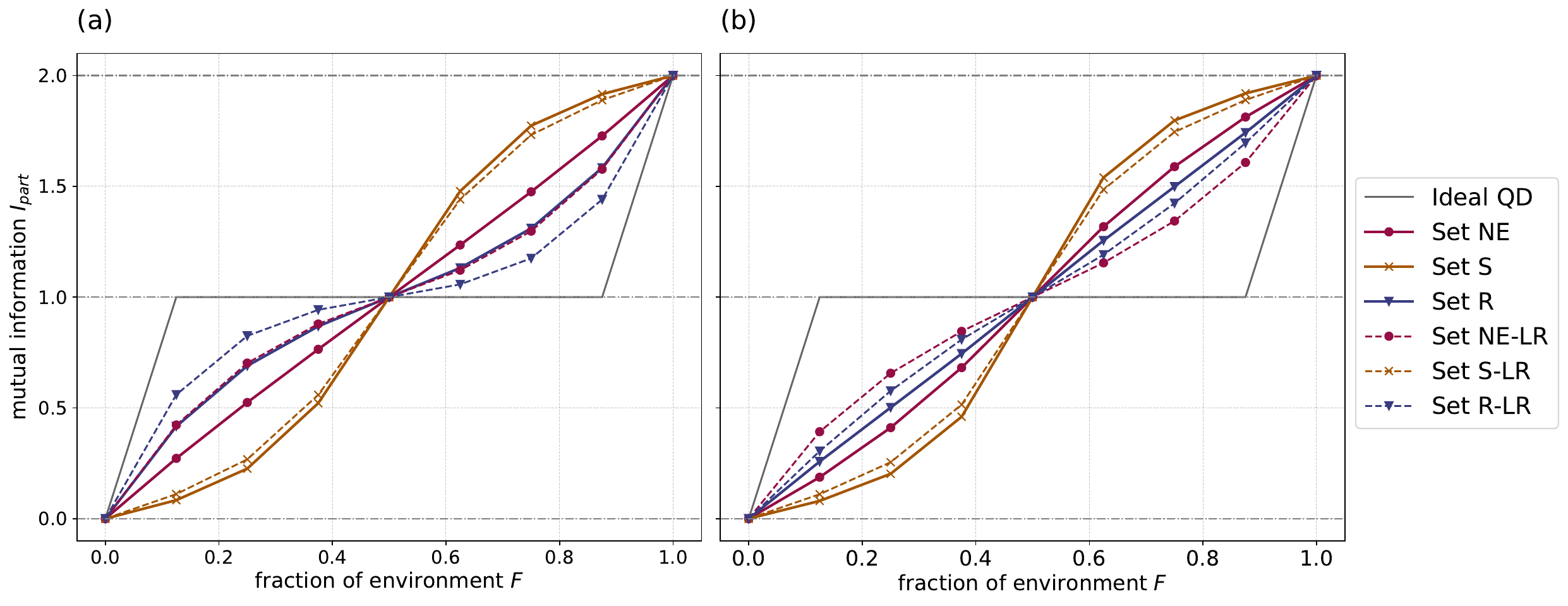}
    \caption{The QMI $I_\text{part}^{\text{F/B}}[\text{I}:\hat D_{1\dots p}]$ for (a) fermions and (b) hard-core bosons. It is calculated using Eq.~\ref{eq:pQMI} as a function of the fraction of the environment $F=p/N_\text{E}$ using particle-based averaging. The six curves in each subfigure show different parameter sets from Tab.~\ref{tab:parametersets}, with $M_\mathrm{S}=16$ and $N_\mathrm{E}=8$. The timesteps at which these curves were calculated are given in the appendix in Table \ref{tab:usedtimes} (they are the same for fermions and bosons). We also show an `Ideal QD' curve indicating what an idealised quantum Darwinism model would predict.}
    \label{fig:pQMI}
\end{figure*}

The particle-based QMI is obtained by tracing out $N_\text{E}-p$ particles, such that the fraction $F$ is defined by $F=p/N_\text{E}$ (see Eq.~\ref{eq:D_Ip}). In Fig.~\ref{fig:pQMI} (a) we plot this QMI $I_\text{part}[\text{I}:\hat D_{1\dots p}]$ as a function of $F$ for six of the parameter sets given in Tab.~\ref{tab:parametersets} (besides Sets PC and PC-LR) in the fermionic case. Figs.~\ref{fig:pQMI} and \ref{fig:sitebased_QMI} also each include a curve indicating what the ideal redundancy plateau of QD would look like in our circumstances. For all of the results in Sec.~\ref{sec:results} we have $M_\text{S}=16$ and $N_\text{E}=8$. (Some analyses of size effects are given in the appendices.)

Note that the times during the dynamics at which each of the curves in Figs.~\ref{fig:pQMI}, \ref{fig:sitebased_QMI}, and \ref{fig:QMI_sites_vs_particles} are calculated is different -- for each different parameter set, the time chosen to plot the QMI as a function of fraction size is determined by two factors: we choose a time at which we observe maximal decoherence ($D(t)\approx 0$ and at a local minimum), and where the QMI has its smallest slope at $F=1/2$ within the evaluated time interval. The timesteps for each of the curves are given in the appendix in Tab.~\ref{tab:usedtimes}.

Looking at the properties of each curve in Fig.~\ref{fig:pQMI} (a) in more detail, consider first Set R for the fermionic case ($W_\text{EE}=0$, $W_\text{IE}=-4$, $i_\text{max}=3$) in Fig.~\ref{fig:pQMI} (a). Here, the QMI as a function of environment fraction size shows clear signs of the characteristic redundancy plateau of quantum Darwinism: with only a small fraction of the environment considered, much of the classical information about the system is present in the environment, and the curve flattens as the fraction approaches half the size of the environment. But it is only when almost all of the environment is considered that the full quantum information in the system become available. The presence of this redundancy curve is why we call this parameter set Set R. It is not too surprising that this set is closer to the Perfect QD curve. A vanishing environment-environment interaction has been identified as a prerequisite for QD \cite{duruisseau23, doucet24}, as otherwise the quantum information is scrambled in the environment, see also \cite{engineer26}. (Though see \cite{21MirkinWisniacki} for a recent work that raises doubts about how harmful environment-environment interactions are for QD.) 

Set R is a short-range case, where the environmental particles interact with the impurity over only $3$ sites. Hence it is also not surprising that this plateau is not as exact as an ideal QD classical plateau. Increasing the range of the interactions to $i_\text{max}=7$ (Set R-LR) leads to a further flattening of the curve and a more pronounced plateau in Fig.~\ref{fig:pQMI} (a) -- the closest to the Perfect QD curve, in fact. This is in line with the observation made for spin systems that fractions of the environment that interact with the subsystem with different interaction strengths (asymmetric environments) lead to a `washing out' of the redundancy curve \cite{chisholm24}.

Contrast this with Set S ($W_\text{EE}=-1$, $W_\text{IE}=-4$, $i_\text{max}=3$) for fermions, which has almost the opposite behaviour in Fig.~\ref{fig:pQMI} (a). With the environment-environment interactions turned on, information about the impurity in the environment is scrambled, similar to the behaviour in Fig.~\ref{fig:decoherence_density}. The implication here is that the information about the impurity position can only be gained from a large fraction of the environment, meaning no redundancy is present. In Fig.~\ref{fig:pQMI} (a), we see that Set S has a steep slope near $F=1/2$. Also in this plot we see that this behaviour is not noticeably changed by increasing the range of the interaction to $i_\text{max}=7$ (Set S-LR).

Next consider Set NE ($W_\text{EE}=-1$, $W_\text{IE}=-1$, $i_\text{max}=3$) for fermions. Its results in Fig.~\ref{fig:pQMI} (a) are an intermediate between the previous two cases -- an almost perfectly linear relationship between fraction size and QMI for periods between revivals (i.e. where $D(t)\approx 0$). During these periods, however, we still expect scrambling interactions to destroy any redundancy plateau, but we find the scrambling to not be as pronounced. This can potentially be explained by the small $d_\text{eff}$ as discussed before -- the system cannot behave like a many-body system and support a many-body effect such as scrambling with such a small effective dimension for the dynamics.

Interestingly, for two of the parameter sets, increasing the range of the interaction does cause a noticeable difference. For fermions, Sets NE-LR ($W_\text{EE}=-1$, $W_\text{IE}=-1$, $i_\text{max}=7$) and R-LR ($W_\text{EE}=-1$, $W_\text{IE}=-1$, $i_\text{max}=7$) have a much more pronounced redundancy plateau than the corresponding short-range sets. (The same is also true for the bosonic case but from a lower starting point, this will be discussed shortly below.) In a sense, the results for fermionic NE-LR appear to defy the usual assumptions about quantum Darwinism, since there are environment-environment interactions but also we have some redundancy. This can perhaps be explained by the fact that this is a low-excitation regime where the interactions do not have as much of a chance to destroy the correlations. But the difference between the short-range and long-range case still requires discussion. It is perhaps indicative of the effects discussed in \cite{chisholm24} -- in the long-range case, the impurity interacts with more of the sites and hence with more environment particles, so the particles can contain more information about the impurity location. This increases the possibility of redundancy and hence increases the QMI. And this effect is least pronounced for Set S versus Set S-LR because of the high excitation, meaning the environment particles have less time to interact with the impurity at the edges. 

On which note, it is now instructive to compare these results for fermions with the corresponding results for hard-core bosons -- see Fig.~\ref{fig:pQMI} (b). For Sets S, S-LR, NE, and NE-LR, we observe very similar curves for the QMI for hard-core bosons as for the fermions (in particular, the NE-LR curves are almost identical).

Only Sets R and R-LR show marked differences between fermions and bosons. While increasing $i_\text{max}$ flattens out the QMIs near $F=1/2$ also in this case, overall the plateaus are much less developed and much closer to being linear in the bosonic case as compared to the fermionic case. The reason why in the fermionic case Set R and Set R-LR show more redundancy properties is discussed in more detail in Appendix \ref{app:spinstatistics}, as the effect also  has implications for the site-based averaging discussed below (see also Fig.~\ref{fig:QMI_sites_vs_particles}). But here we can say that the varying excitation of the system for different Hamiltonian parameters plays a role. For Sets S and S-LR, which are the energetically most excited ones compared to the other sets, statistics plays a minor role in the overall dynamics due to the comparatively large excitation, while in the NE cases, which are very weakly excited and the effective dimension is very small, the available state space for the particles to explore is limited. Both effects lead to smaller sensitivity to particle statistics in the QMI, and thus only Sets R and R-LR are in the appropriate energy range to demonstrate these effects.

Our analysis shows that, especially at intermediate excitation energies pertinent to our present study, there are marked differences between fermions and bosons when it comes to their ability to store information about the impurity, and this difference is most pronounced in precisely the Hamiltonian parameter sets identified as crucial for QD. Hence we can say that, while for a fermionic environment information is being broadcast and stored in a redundant way, this information is partly lost in bosonic environments.

\subsubsection{The mutual information with site tracing}\label{subsubsec:site_entang}

As discussed before, the way in which a `fraction' of the environment is defined makes an important difference to the structure of the QMI curves. In this subsection we will study how the QMI is structured when fractions are defined based on the number of sites considered, rather than the number of particles. Since the sites are in general different from one another, this process will be much more subtle.

Here we will focus more on the parameter sets with long-range (LR) interaction only, as all relevant features in the QMIs are more pronounced in this case. First, recall that $X$ labels an ordered tuple of $m$ sites $X=\{j_x\}_{x=1}^{m}$, e.g.~$X=\{1,2,3\}$ or $X=\{4, 8, 14, 15\}$. Then the site-based reduced density matrix $\hat{\rho}_X$ is obtained by tracing over the complement of $X$ as in Eq.~\ref{eq:rho_S} and then further tracing over the impurity. This then allows us to calculate the site-based QMI for a specific tuple of sites using Eq.~\ref{eq:sQMI}. We can then average the QMI over all possible tuples of a certain cardinality $|X|=m$ using Eq.~\ref{eq:sQMI_aver}. This is in general an extremely expensive process as there can be many possible tuples per cardinality. The size of the fraction of the environment, in this case, is simply $F=|X|/M_S$.

\begin{figure}[t]
    \centering
    \includegraphics[width=\linewidth]{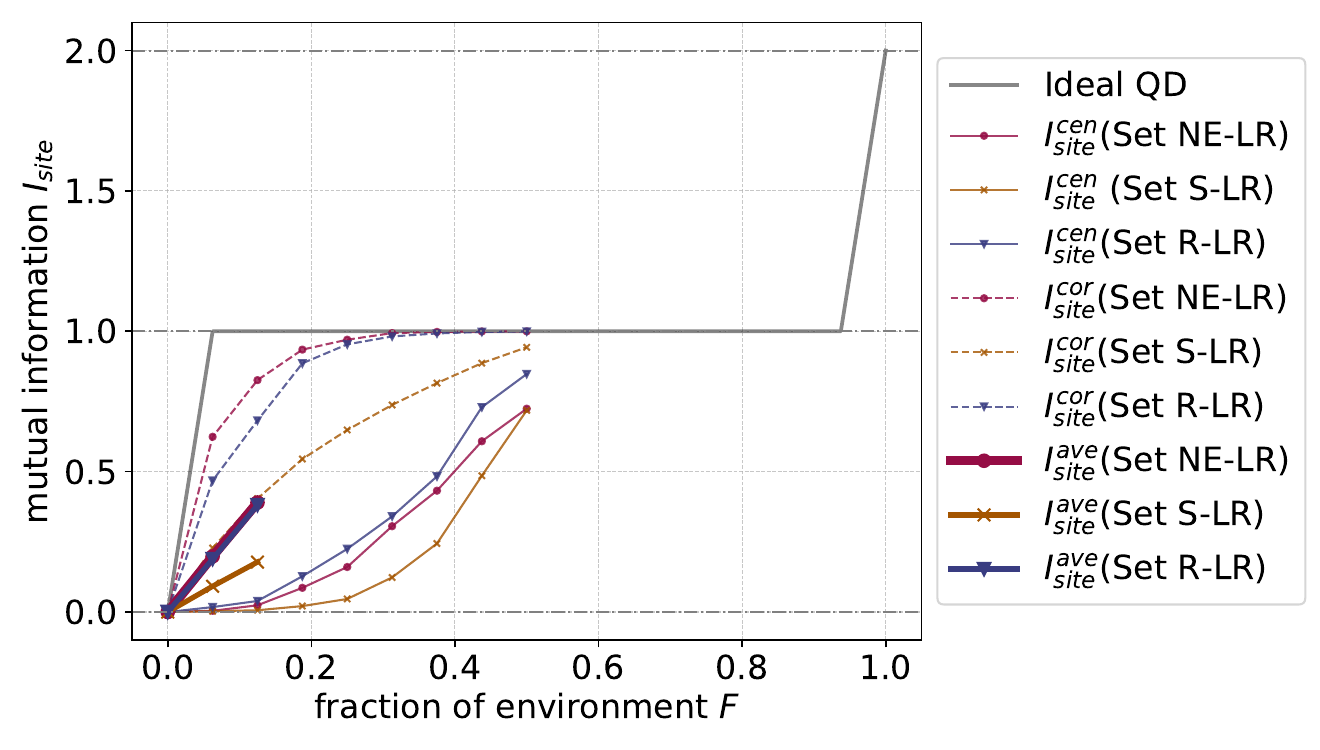}
    \caption{The QMI based on traces over sites $I_\text{site}$ (Eq.~\ref{eq:sQMI}) for $M_\text{S}=16$ and $N_\text{E}=8$. (Fermions were used in this calculation but the result is identical for bosons.) Each curve represents a certain set of parameters from Tab.~\ref{tab:parametersets}, and different choices of $X$, the set of sites to average over. The set is either filled from the left and right edges of the system (labelled `corner', $I^\text{cor}_\text{site}$), or from the centre of the system (labelled `centre', $I^\text{cen}_\text{site}$), or we average over all possible choices for a given $|X|$: $I^\text{ave}_\text{site}$ from Eq.~\ref{eq:sQMI_aver}. As before, we include an idealised QD curve. For each curve, the timestep at which the calculation was performed is listed in the appendix in Tab.~\ref{tab:usedtimes}.}
    \label{fig:sitebased_QMI}
\end{figure}

Different sites will in general contain different amounts of information about the impurity's position. Broadly speaking, sites closer to the edges will contain more information, for instance. And so it is worth exploring how different choices of the set $X$ for a given $F$ affects the QMI. We start by choosing the sites that maximise the plateau-like behaviour of QMI (such that it is maximal for $F<1/2$), which means choosing sites from the edges inwards. Hence, we choose for $|X|=1$ the leftmost site $\{1\}$, for $|X|=2$ we append to this tuple the rightmost site $\{1,16\}$, for $|X|=3$ we append the site neighbouring the leftmost site $\{1,2,16\}$, and so on. We label the corresponding QMIs $I^\text{cor}_\text{site}[\text{I}:\hat \rho_{X}]$ to indicate that the sets $X$ grow starting from the corners. In Fig.~\ref{fig:sitebased_QMI}, we plot this quantity for the three long-range parameter sets: NE-LR, S-LR, and R-LR.

In addition to $I^\text{cor}_\text{site}[\text{I}:\hat \rho_{X}]$, for each of the three parameter sets Fig.~\ref{fig:sitebased_QMI} also shows two more choices for which sites to average over for a given cardinality. First, we consider the case where we start in the centre, so with $\{8\}$ for $|X|=1$, $\{8,9\}$ for $|X|=2$, $\{7,8,9\}$ for $|X|=3$ and so on. This centre-based QMI is labelled $I^\text{cen}_\text{site}[\text{I}:\hat \rho_{X}]$. Then, we also show the case where we average over all possible tuples of sites with Eq.~\ref{eq:sQMI_aver}: $I^\text{ave}_\text{site}[\text{I}:\hat \rho_{X}]=\langle I_\text{site}[\text{I}:\hat \rho_X]\rangle_{|X|}$.

Note that for this averaged case, the computational difficulty of taking the average meant that we only calculated a small number of points on the curve. While for $|X|$ of up to $|X|=3$ it is still numerically feasible, for $|X|>3$ obtaining all QMIs starts to become prohibitive. For example, $|X|=4$ amounts to an average over $\binom{16}{4}=1820$ different tuples $X$, and for each of these tuples, the reduced density matrices require traces over the entire configuration space of $\text{dim}(\mathcal{H})=25740$. Fortunately, it is enough to evaluate the QMI for up to $|X|=3$ to observe all relevant trends. And for each of the parameter sets and averaging choices, we plot the QMI only for $F\leq 1/2$, since larger $|X|$ are increasingly demanding computationally.

For the parameter sets R-LR and NE-LR (which both showed redundancy in the fermionic case), Fig.~\ref{fig:sitebased_QMI} also shows pronounced redundancy plateaus for the site-based QMI with corner-based partial tracing. Interestingly, the plateau is more pronounced for the NE-LR case (the case with environment-environment interactions) compared to the R-LR case (which lacks environment-environment interactions). Based on the choice for how we construct $X$ (starting from the corners or starting from centre of the system), we do not expect the QMI curve to be symmetric with respect to $F=1/2$. (But we do expect the averaged $\braket{I_\text{site}[\text{I}:\hat \rho_X]}_{|X|}$ to be symmetric as it is for the QMI based on particle traces.)

In the site-based case with corner averaging, we find that the QMI is smaller for Set R-LR than it is for Set NE-LR. This is most likely due to the strong spatial particle fluctuations present in the former of those. (Compare also the conditioned density fluctuations in Fig.~\ref{fig:decoherence_density} for the sets NE and R, which show the same trend in this respect.) The reason for this is that entropies of the reduced density matrices conditioned on the impurity, $S(\hat \rho_{\text{I},X})$, are larger than the entropies $S(\hat \rho_{X})$ subject to the trace over the impurity. The trace over the impurity smooths out the fluctuations and thus decreases the entropy. The site based QMIs are sensitive to these spatial fluctuations, while the QMIs based on particle traces are sensitive to the eigenvalues of the particle reduced density matrices which pertain to states that can be spread over the entire system.

For the centre-averaging $I^\text{cen}_\text{site}[\text{I}:\hat \rho_{X}]$, small sets $X$ tend to have no information about the impurity. The environment particles only have information about the impurity when they cluster on sites near the impurity positions at the ends of the chain, and so very little information about the impurity can be stored in the centre. In Fig.~\ref{fig:sitebased_QMI}, we see for each parameter set the opposite behaviour for the centre-based averaging as we do for the corner-based one. The Sets R-LR and NE-LR, which show strong plateaus for the corner-based averaging, show almost perfect anti-redundancy for the centre-based averaging, and Set S-LR, which showed less of a plateau for the corner-based case, shows even less information-containing behaviour in the centre-based case. In other words, for each parameter set in Fig.~\ref{fig:sitebased_QMI} we see that $I_\text{site}^\text{cen}$ only starts to increase substantially when $|X|\gtrsim 3$, as before that, all of the sites are too close to the centre to contain any information about the impurity's location

\begin{figure}[t]
    \centering
    \includegraphics[width=1\linewidth]{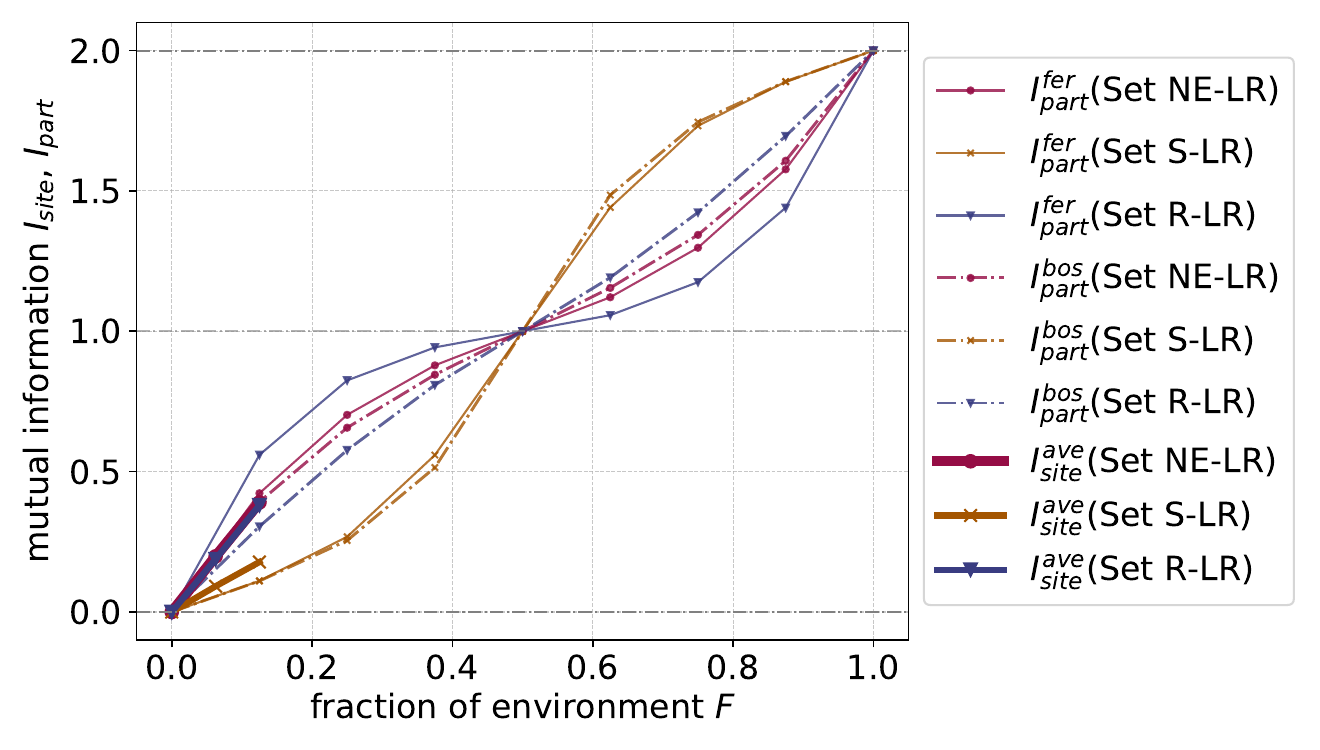}
    \caption{Comparison between the averaged QMI based on traces over fermions and bosons $I_\text{part}^{\text{F/B}}[\text{I}:\hat D_{1\dots p}]$ (as in Fig.~\ref{fig:pQMI}), and the QMI based on traces over sites $\braket{I_\text{site}[\text{I}:\hat \rho_X]}_{|X|}$ (as in Fig.~\ref{fig:sitebased_QMI}). For two of the parameter sets, the three different QMIs agree with each other. 
    }
    \label{fig:QMI_sites_vs_particles}
\end{figure}

Overall, the results for the QMI based on site traces are consistent with the previous analyses of QMIs based on particle traces. In Fig.~\ref{fig:QMI_sites_vs_particles} we compare the averaged site-based QMI $I^\text{ave}_\text{site}[\text{I}:\hat \rho_{X}]$ to the fermionic and bosonic particle-based QMIs $I_\text{part}^{\text{F/B}}[\text{I}:\hat D_{1\dots N_{\text{E}}-p}]$.

Even though we use much fewer data points for the site-based averaging, it is clear from Fig.~\ref{fig:QMI_sites_vs_particles} that they follow the same trends as the  particle-based QMIs for each parameter set. For instance, we see that no redundancy appears for the parameter Set S-LR if averaged over all $X$, similar to how no redundancy is present for the fermionic or bosonic S-LR sets. Likewise, Set NE-LR appears to be almost identical for the fermion, boson, and averaged site-based cases, with hints of a diminished redundancy plateau.

The curve for $I^\text{ave}_\text{site}[\text{I}:\hat \rho_{X}]$ indicates that there is practically no plateau near $F=1/2$ in this case. It seems that averaging over all sets $X$ washes out any information stored close to the impurity. The origin of this might again be the particle fluctuations across the system (which one can think of as `noise'), which are larger for the conditioned $\hat \rho_{\text{I},X}$ as compared to the traced $\hat \rho_{X}$. In this regard, the impurity has a similar effect on increasing the entropy for $S(\hat \rho_{\text{I},X})$ compared to $S(\hat \rho_{X})<S(\hat \rho_{\text{I},X})$, as in the bosonic case where the impurity disturbs the bosonic bunching and $S(\hat D_{1\dots p})<S(\hat D_{\text{I},1\dots p})$ (see Appendix.~\ref{app:spinstatistics}).

The only parameter set where there is any deviation in Fig.~\ref{fig:QMI_sites_vs_particles} is Set R-LR. Now, the site-based averaging case appears to behave the same as the bosonic case, while the fermionic case behaves very differently. The reason for this deviation in Fig.~\ref{fig:QMI_sites_vs_particles} is quite subtle, and is discussed in Appendix \ref{app:spinstatistics}. Though here we point out that previous works have shown that fermions can have an advantage over bosons for transferring information \cite{07SenBrussLewenstein}, and can have different entropic and QMI properties \cite{09HaqueZozulyaSchoutens, 26TengXuYang}.
\section{Conclusions}
\label{sec:conclusions}

In quantum theory, concepts such as measurements, decoherence, and the quantum-to-classical transition are often described interchangeably. And over the years many different notions of classicality have been developed that all capture various key differences between the quantum regime and our intuitive understanding of classical reality. In the quantum Darwinism literature, the core indicator of classicality is the redundant encoding of information from the pointer basis of a system into its surroundings, leading to the emergence of persistent, stable, objective measurement outcomes. Many works have studied the circumstances under which we observe properties of the emergence of classicality like decoherence and QD's well-known redundancy plateau. Here, we contribute to that literature with a realistic, tunable many-body (exact) numerical model, where the circumstances under which the dynamics drives a system to express features of QD are clear and measurable.

In particular, we have numerically analysed the dynamics of a 1D system of site-hopping particles, which we call an environment, influenced by the presence of a quantum system, which we call an impurity. For the impurity, initially in a superposition between being located at two maximally separated sites, we investigated under which conditions the environment particles act to record the location of the impurity, analogously to how an apparatus measures a system in the von Neumann model of measurement.

We investigated the dynamics of the impurity-environment system under a variety of parameter regimes. With the impurity kept static, we varied how much and over what range it interacted with the environment particles, whether or not the environment particles interacted with each other, and the exchange symmetry that the particles involved possessed. Decoherence was straightforward to observe in this study: we saw for most parameter regimes a mixed state of the environment that clearly decomposed into one state commensurate with the environment having `detected' the impurity on the left, and another state commensurate with the right. Hence, the location of the impurity served as a proper pointer basis for decoherence, which allowed us to use this basis to study QD properties of the system. The presence of a pointer basis was guaranteed through the structure of the impurity-environment interaction Hamiltonian \cite{doucet24}, but the open questions we studied here concerned how the decoherence process affected the environment itself. 
Crucially, in this work we addressed the question of how to evaluate the information about the system that is stored in the environment when the particles in the environment are indistinguishable. This was done by analysing the QD properties of the impurity-environment system when the environment was made of either fermions or hard-core bosons. In order to quantify the amount of information the environment stored in each case, we evaluated the quantum mutual information as a function of an increasing fraction of the environment considered. This required us to devise ways to meaningfully define environment fractions when the environment was composed of indistinguishable particles. Additionally, we addressed the question of how to average over particles versus how to average over orbitals (with the latter characterised as lattice sites in our model). We then showed how the QMI for a fraction of the environment can be obtained both with particle-based partial tracing and site-based partial tracing. Notably, by doing this we found that the QMI based on particle traces was sensitive to the particle statistics, while the QMI based on site traces was not.

We considered the dynamics of the system for different choices of Hamiltonian parameters, corresponding to increased or decreased impurity-environment interaction and increased or decreased environment-environment interaction. An ideal manifestation of QD has a `redundancy plateau' in the plot of the QMI as a function of the size of the environment fraction. In our case, however, we did not observe the plateau fully emerge even in cases with no environment-environment interactions. We found that more plateau-like behaviour emerged for some parameter sets (including some with environment-environment interactions), and interestingly, in some cases the plateau was more pronounced for fermions than for bosons. Additionally, in accordance with findings in \cite{chisholm24}, we saw that in a site-based picture, the choice of which sites to average over severely impacted whether a plateau emerged or not.

Typically, the key features of the theory of quantum Darwinism emerge when there is no interaction between environment particles. In our work, we did observe the appearance of a redundancy plateau for certain cases involving an interacting environment. Previously these interactions were assumed to prevent the redundant storage of information due to scrambling (though recent works have called this into question \cite{21MirkinWisniacki}). However, we show that for properly chosen energy scales and weak excitation of the total system, interactions between environment particles regularises the overall dynamics, meaning that we observe a regular back-and-forth motion of the particles, pronounced decoherence effects whenever the environmental particles come close to the impurity, and large revivals of coherence for certain periods of time. It is telling that in this case, all the QMI curves (based on particle traces for bosons and fermions, as well as the QMI based on site traces) agreed with each other. 

As alluded to in the Introduction, for a model of QD to function as a proper representation of a measurement, we require a way for a particle's microstate to create a macroscopic change in the environment. This can be achieved by engineering an environment that behaves like a detector with an active amplification mechanism, which expends thermodynamic resources to amplify signal from the system. It is an open question whether the model we develop here can be adapted in such a way, potentially to align with theoretical models of actual detector apparatuses. (A discussion of phase transitions may also be appropriate in this context.) Furthermore if we wished to study longer timescales and thermalisation, we would need to modify our model to allow the impurity to move around, which would drastically alter the results presented here, but also make the simulation serve much less well as a model of the measurement process.

Given that our numerical model is inspired by experiments involving ultracold atom simulators in optical lattices, we assert that our work opens up the research on fundamental aspects of quantum measurement dynamics to the platform of ultracold atoms. High tunability is required for connections between numerical models and experiments, in particular with respect to the strengths of the various interactions, the properties of the impurity atoms, and the exchange symmetry of the particles. Much of this tunability can already be realised experimentally, and we emulate that with our model. Numerical studies such as this one rely on exact diagonalisation, which restricts the size and dimensionality of the system being studied. Despite the problem of the QMI being somewhat difficult to handle as an experimental observable, there are benefits to using lab-based ultracold atom simulators to study QD effects. Such lab setups could enable the study of larger and higher-dimensional systems, in which redundant storage of information could be even more pronounced.

\begin{acknowledgements}
\noindent  This research was funded by the Austrian Science Fund (FWF) Grant Nos.~10.55776/COE1, 10.55776/P35539, 
10.55776/ESP7464924, and 10.55776/I6949. M.~H.~acknowledges funding from the European Research Council (Consolidator grant ‘Cocoquest’ 101043705).
 Calculations were performed on the Vienna Scientific Cluster (VSC4 and VSC5). This publication was made possible through the support of Grant No.~62423 from the John Templeton Foundation. The opinions expressed in this publication are those of the authors and do not necessarily reflect the views of the John Templeton Foundation. We thank Florian Grüner, Yuri Minoguchi, Elias Pescoller, Jan Neuser, Danial Chughtai and Alberto Rolandi for helpful discussions. AI acknowledgement: ChatGPT was used as a coding assistant for Python syntax throughout the project (with the latest models from 2024-2026). All code was checked by the authors as it was being written and all outputs were verified at each step.
\end{acknowledgements}

\appendix
\section{Parameter sets}
\label{app:params}
We present here for completeness all the parameters used for the different simulations discussed in the main text, see Tab.~\ref{tab:parameterdetails}. In the main text, we keep the number $M_\text{S}=16$ and $N_\text{E}=8$. In later parts of the Appendix we will analyse the effect of system size. For simplicity, we have chosen how the interaction between the impurity and the environmental particles scales with distance to be (close to) linear. Small deviations from linearity do not make a difference in our results. The spatial decay of the interactions has an influence on the effective number of environmental particles that interact with the impurity and thus on how flat the QMI curve is near $F=1/2$.

\begin{table}[h]
\begin{center}
\begin{tabular}{r|c|c|c|c|c|c|c|c|c}
     Set  & $W_{\mathrm{IE}}$ & $W_\mathrm{EE}$ & $s_1$ & $s_2$ &$s_3$ &$s_4$ &$s_5$ &$s_6$ &$s_7$ \\
     \hline
     PC&-1 & 0 & $4/5$ & $1/2$ & $3/10$ & 0 & 0 & 0 & 0\\
     NE&-1 & -1 & $4/5$ & $1/2$ & $3/10$ & 0 & 0 & 0 & 0\\
     S&-4 & -1 & $4/5$ & $1/2$ & $3/10$ & 0 & 0 & 0 & 0\\
     R&-4 & 0 & $4/5$ & $1/2$ & $3/10$ & 0 & 0 & 0 & 0\\
     NE-LR&-1 & -1 & $7/8$& $6/8$&$5/8$& $4/8$& $3/8$& $2/8$& $1/8$  \\
     S-LR&-4 & -1  & $7/8$& $6/8$&$5/8$& $4/8$& $3/8$& $2/8$& $1/8$ \\
     R-LR&-4 & 0 &$7/8$& $6/8$&$5/8$& $4/8$& $3/8$& $2/8$& $1/8$ 
     \label{tab:deff_tdec}     
\end{tabular}
     \caption{Full parameter sets for the systems discussed in the main text.}
     \label{tab:parameterdetails}
\end{center}
\end{table}

In Tab.~\ref{tab:usedtimes}, we list the timesteps used to calculate each of the quantum mutual information plots in the main text. For each given parameter set, a time was chosen at which $D(t)$ was at a minimum (and sufficiently long after equilibration was initially reached), which for the parameter sets of interest always meant that it was approximately zero. This was to ensure that the impurity-environment system was as decohered as possible, to give the best chance of observing redundancy in the QMI curve. More analysis of this can be found in Appendix.~\ref{app:decor_dyn}. The values in the table were found by calculating 400 time steps, each being of size $\Delta t = 0.2$ in the energy units set by $J_E=1$. For each given parameter set, the same time was used to calculate $I^\text{F}_\text{part}$, $I^\text{B}_\text{part}$, $I^\text{cor}_\text{site}$, $I^\text{cen}_\text{site}$, and $I^\text{ave}_\text{site}$.

\begin{table}[t]
    \centering
    \begin{tabular}{r|c}
        Set & $t_{\mathrm{QMI}}$\\
        \hline
         Set NE & 67.2 \\
         Set S &  69.4\\
         Set R &  39.2\\
         Set NE-LR &11.8  \\
         Set S-LR & 37.0  \\
         Set R-LR &  62.0
    \end{tabular}
    \caption{The timesteps $\tau_{\mathrm{QMI}}$ at which the QMIs were calculate for each parameter set, obtained by minimising $D(t)$ (in each case it is the same for fermion-, boson-, and site-based tracing). }
    \vspace{-5pt}
    \label{tab:usedtimes}
\end{table}

\section{How exchange symmetry affects the QMI }\label{app:spinstatistics}

\begin{figure*}
    \centering
    \includegraphics[width=\linewidth]{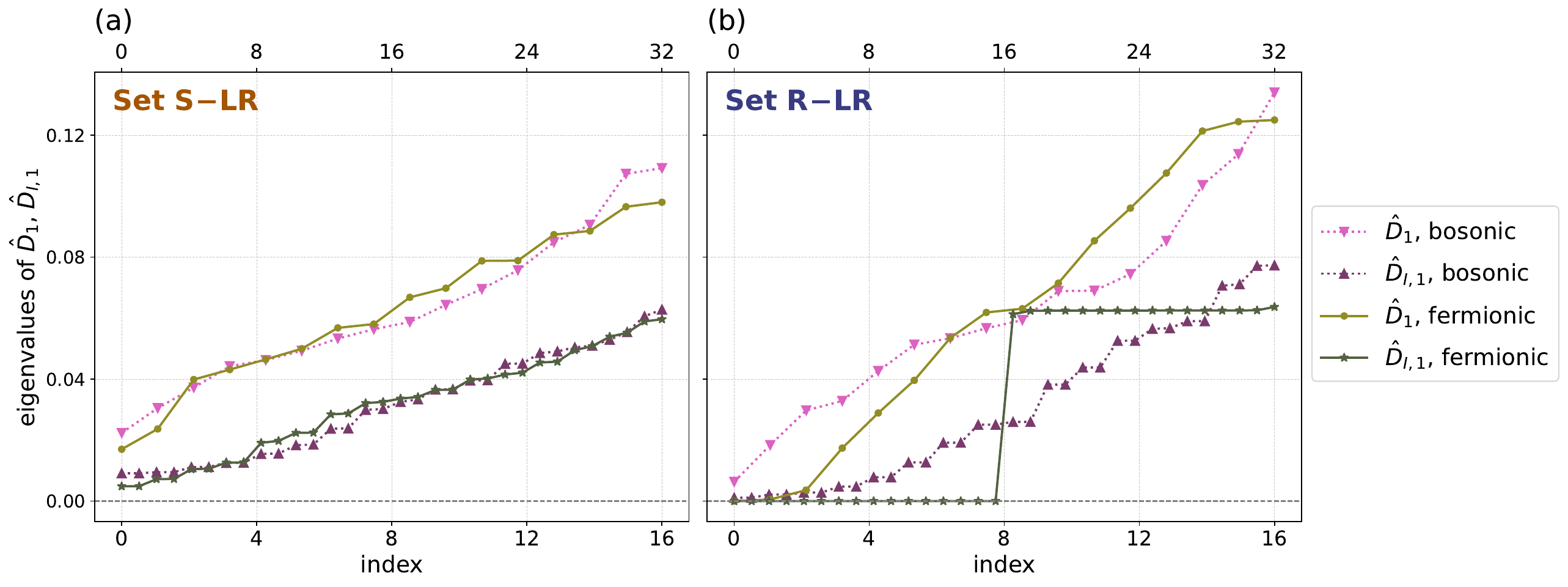}
    \caption{The eigenvalues of $\hat D_{\text{I},1}$ and $\hat D_{1}$ from Eq.~\ref{eq:D_Ip} for (a) Set S-LR and (b) Set R-LR, for both fermions and bosons (for $M_\text{S}=16$ and $N_\text{E}=8$). $\hat D_{\text{I},1}$ has 32 eigenvalues, and the indices labelling each of its eigenvalues are listed on the top. $\hat D_{1}$ has 16 eigenvalues, and its eigenvalue indices are listed on the bottom. For $\hat D_{1}$, the 16 indices correspond to the 16 sites, but for $\hat D_{\text{I},1}$, the 32 indices correspond to the 16 sites conditioned on the impurity being left or right.}
    \label{fig:fermboseigvals}
\end{figure*}

Here we explain in more detail our observation that for certain parameter choices (Sets R and R-LR), the fermionic case appears to show notably more redundancy than the bosonic one, manifesting in a QMI curve that is closer to the ideal QD plateau in Figs.~\ref{fig:pQMI} and \ref{fig:QMI_sites_vs_particles}. Part of the explanation for this appears to lie in the eigenvalues of the particle-based reduced density matrices (RDMs). In principle this should be an effect that is most visible for the larger RDMs, but we already see the effect in the two smallest ones: $\hat D_{\text{I},1}$ and $\hat D_{1}$ (the former coming from Eq.~\ref{eq:D_Ip} and the latter obtained by tracing out the impurity from the former), which are the simplest to analyse. 
In Fig.~\ref{fig:fermboseigvals}, we plot the eigenvalues of these two matrices in both the fermionic and bosonic cases for two parameter sets (Set S-LR and Set R-LR). The most striking feature of this plot is a `Fermi edge' in the eigenvalue distribution of $\hat D_{\text{I},1}$ for Set R-LR (Fig.~\ref{fig:fermboseigvals} (b)), especially notable as it is not present in $\hat D_{1}$ for that parameter set, or in either fermionic RDM for Set S-LR (Fig.~\ref{fig:fermboseigvals} (a)).

Thanks to decoherence, $\hat D_{\text{I},1}$ is neatly split into a block-diagonal structure with blocks conditioned on the impurity being on the left or the right. The eigenvalues can be thought of as representing the occupation of the `natural orbitals' of the system (not to be confused with the site-based orbitals). It appears as though, for Set R-LR (Fig.~\ref{fig:fermboseigvals} (b)), fermionic anti-bunching causes the eight left- and right-most natural orbitals to become perfectly saturated, which implies that they leave the remaining $16$ ($8$ each for the left- and right-conditioned blocks) perfectly unfilled. (This effect is also visible in plots of the eigenvectors, but we have chosen not to include these plots here.) This explains the structure of the fermionic curve for $\hat D_{\text{I},1}$ for Set R-LR (and a similar effect is seen for Set R, Set NE and Set NE-LR). But $\hat D_{1}$ is calculated by averaging over the left- and right-conditioned cases, which washes out the Fermi edge structure. 

This structure is not present for the bosons because they are not compelled to spread out among the natural orbitals. Even though the bosons are hard-core and thus cannot doubly occupy a site, they still observe bosonic clustering (or `bunching') in the space of the natural orbitals. Including the impurity then disturbs this bunching effect: because the eigenvalues of $\hat D_{\text{I},1}$ are pairwise degenerate, the impurity shifts the larger eigenvalues of $\hat D_{1}$ more by `fragmenting' them among left and right in $\hat D_{\text{I},1}$. The effect of this is that the distribution of eigenvalues for $\hat D_{\text{I},1}$ is flatter than the distribution for $\hat D_{1}$ (for any bosonic case), but there is no substantial difference in the shape of the two distributions as there is in the fermionic case.  

The connection to the redundancy properties is via the QMI, which itself depends on the difference in entropy between $\hat D_{\text{I},1}$'s and $\hat D_{1}$'s eigenvalues. 

One can calculate the entropy associated with the eigenvalue distribution $\{\lambda_i\}_i$ as simply $-\sum\lambda_i{\rm{log}_2}\lambda_i$. Doing this for the two RDM matrices $\hat D_{\text{I},1}$ and $\hat D_{1}$ (for the fermionic and bosonic S-LR and R-LR cases) leads to the entropies $S(\hat D_{\text{I},1})$ and $S(\hat D_1)$ respectively in Tab.~\ref{tab:entropies} (we also include the entropies for Set NE-LR). In that Table we can also see the differences between the entropies, confirming that for Set S-LR the difference in entropies is the same for fermions and bosons, but for Set R-LR they are notably different.

\begin{table}[]
    \centering
    \begin{tabular}{r|c|c|c}
        Set & $S(\hat D_{\text{I},1})$ & $S(\hat D_1)$ & $S(\hat D_{\text{I},1})-S(\hat D_1)$\\
        \hline
         Set S-LR (fer) & 4.77 & 3.88 & 0.89 \\
         Set S-LR (bos) & 4.77 & 3.88 & 0.89\\
         Set R-LR (fer) & 4.00 & 3.56 & 0.44 \\
         Set R-LR (bos) & 4.48 & 3.78 & 0.70 \\
         Set NE-LR (fer) & 4.08 & 3.50 & 0.58 \\
         Set NE-LR (bos) & 4.06 & 3.44 & 0.62
    \end{tabular}
    \vspace{-5pt}
    \caption{Entropies of the eigenvalue distributions for $\hat D_{\text{I},1}$ ($S(\hat D_{\text{I},1})$) and $\hat D_{1}$ ($S(\hat D_1)$), and their differences, for each of the six long-range parameter sets considered in the main text, calculated at the times $t_{\mathrm{QMI}}$ in Tab.~\ref{tab:usedtimes}. (Note that the S-LR values are only the same up to three significant figures.)}
    \vspace{-10pt}
    \label{tab:entropies}
\end{table}

Generally speaking, we are concerned with the ways in which the impurity disrupts the fermionic anti-bunching and bosonic bunching when it is added to the system. It appears as though in the fermionic case, the impurity enhances the anti-bunching effect, since it is characterised by the Fermi edge, which is not present in $\hat D_1$ (meaning that anti-bunching is less pronounced in $\hat D_1$). This seems counter-intuitive, as it suggests that tracing over the impurity gives the fermions more states to spread out into. Then in the bosonic case, the impurity disturbs bunching, since it causes pairwise degeneracy in the states. Hence, bunching is more pronounced in $\hat D_1$.
We observe that this phenomenon causes the difference $S(\hat D_{\text{I},1})-S(\hat D_1)$ to be larger for bosons than for fermions. This must be related to the fact that the presence of the impurity affects bosonic bunching differently to how it affects fermionic anti-bunching. But precisely why this happens remains an open question. It is not precisely clear why an effect that leads to $\hat D_{\text{I},1}$ being flatter than $\hat D_1$ for both fermions and bosons leads to this entropy difference being different for each exchange symmetry, and we leave this question open here.


\section{Dynamics during decoherence}\label{app:decor_dyn} 
Here we provide further numerical data underpinning the results in the main text, specifically to assess how decohered the system is at the times we calculate the QMI. We focus here exclusively on fermions and the QMI based on the particle-reduced density matrices.

First, we show for one example (Set NE, see Tab.~\ref{tab:parametersets}) the behaviour of the eigenvalues of the impurity reduced density matrix (IRDM) $\hat \rho_\text{I}$ -- the density matrix left over when all environment particles have been traced out. Due to the impurity being constrained to stay in either the leftmost or rightmost site, its Hilbert space dimension is $\text{dim}(\mathcal H_\text{I})=2$. Recall also that the impurity starts off in a pure state given by the superposition between the left and the right site leading to eigenvalues of $\hat \rho_\text{I}$ given by $\lambda_1=1$ and $\lambda_2=0$. 

\begin{figure}[h]
    \centering
    \includegraphics[width=1\linewidth]{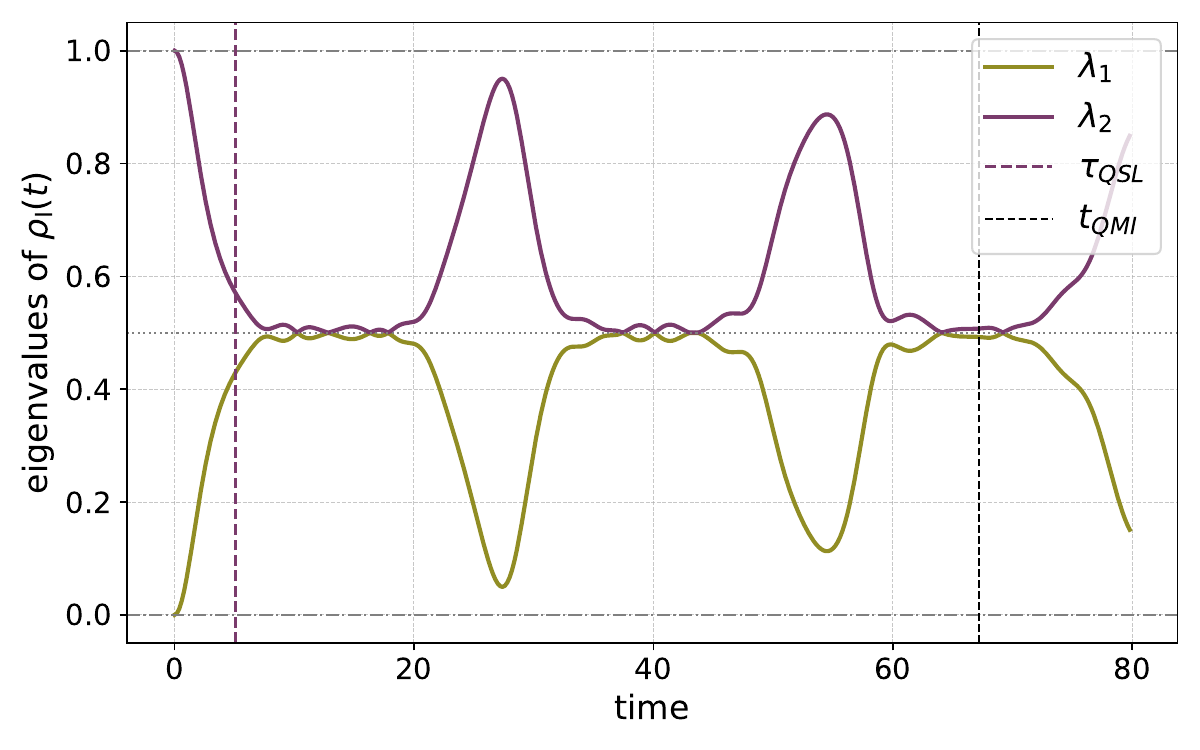}
    \caption{The eigenvalues $\lambda_{1,2}$ of the reduced density matrix of the impurity $\hat{\rho}_\text{I}(t)$ for fermionic Set NE, as a function of time. $\text{dim}(\mathcal{H}_\text{I})=2$, $M_\text{S}=16$, $N_\text{E}=8$. The first vertical line indicates the quantum speed limit $\tau_{\text{QSL}}$ for this parameter set, and the second vertical line indicates the time value used to calculate the QMIs for this parameter set in Fig.~\ref{fig:pQMI}. } 
    \label{fig:impurity_rdm}
\end{figure}
In Fig.~\ref{fig:impurity_rdm} we show the two eigenvalues of this IRDM as a function of time for Set NE (the IRDM is not impacted by exchange symmetry). During the interaction with the environment, as soon as the decoherence parameter reaches $D(t)\approx 0$, we observe that $\hat \rho_\text{I}$ is effectively perfectly mixed state with $\lambda_{1}\approx \lambda_2\approx 1/2$. The dynamics of the eigenvalues aligns well with that of $D(t)$ (see Fig.~\ref{fig:decoherence_density} (a)). For example, the large revivals in $D(t)$ are reflected in revivals at the same time in $\lambda_1$ and $\lambda_2$. This is further justification of the assertion that the system is decohered when $D(t)\approx 0$, even for a parameter set such as this one with large revivals of the coherence.

\begin{figure}[t]
    \centering
    \includegraphics[width=1\linewidth]{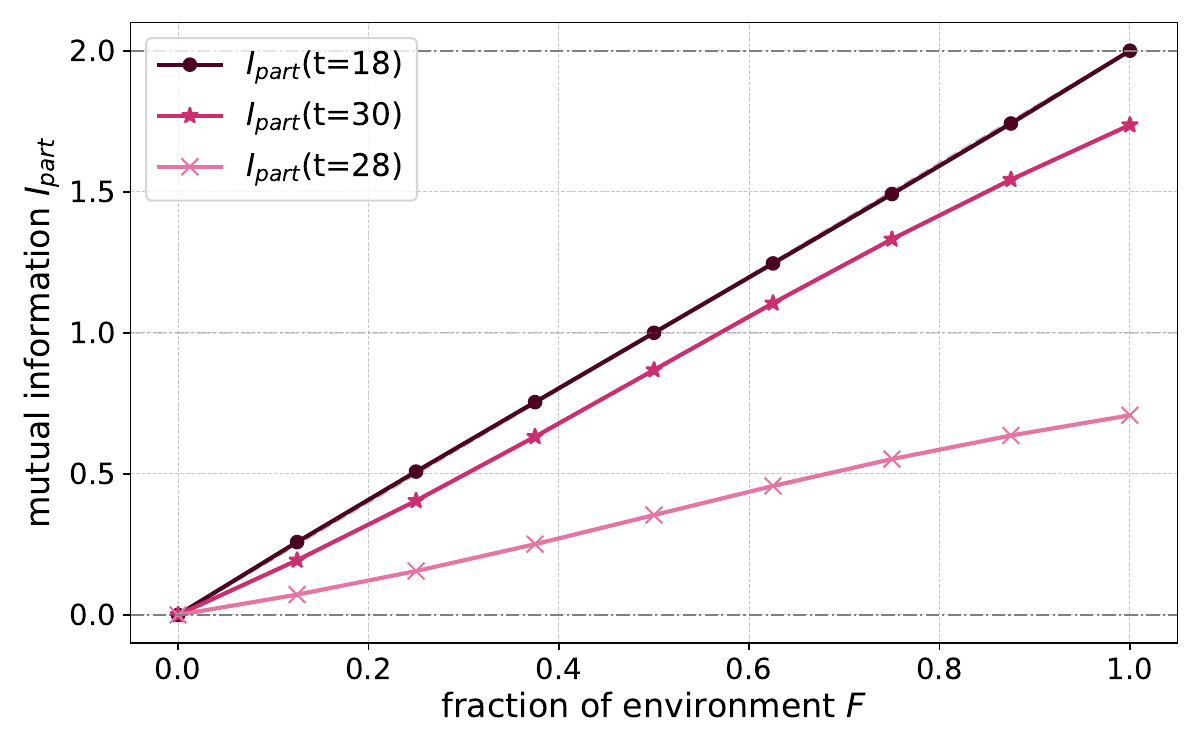}
    \caption{QMI based on particle traces $I_\text{part}^{\text{F}}[\text{I}:\hat D_{1\dots p}]$ for fermionic Set NE for three different times close to the first revival ($M_\mathrm{S}=16$ and $N_\mathrm{E}=8$). Compare to the QMI values in Fig.~\ref{fig:pQMI} (a), and compare the times to those shown in Fig.~\ref{fig:decoherence_density} (a).}
    \label{fig:QMI_revival}
\end{figure}

\begin{figure}[b]
    \centering
    \includegraphics[width=\linewidth]{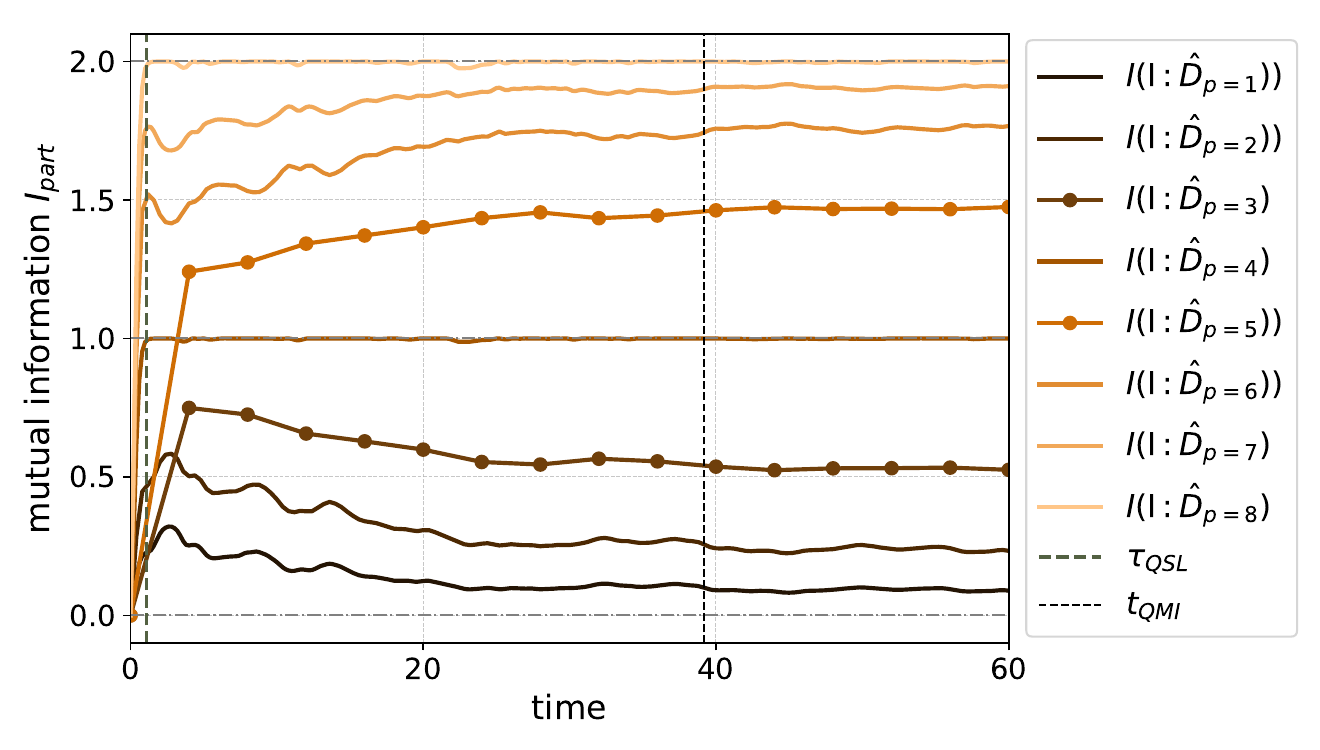}
    \caption{The QMI $I_\text{part}[\text{I}:\hat D_{1\dots p}]$ for fermionic Set S as a function of time ($M_\mathrm{S}=16$ and $N_\mathrm{E}=8$). Each line corresponds to the QMI between the impurity and the given number of environmental particles $p$. For $p=3$, and $p=5$ the QMI has been evaluated only at few time steps indicated by the dots in the figure due to the large numerical effort involved in obtaining the QMI from these large reduced density matrices. The first vertical line indicates the quantum speed limit $\tau_{\text{QSL}}$ for this parameter set, and the second vertical line indicates the time value used to calculate the QMIs for this parameter set in Fig.~\ref{fig:pQMI}.}
    \label{fig:QMI_SetB_time}
\end{figure}

Next, we further justify our assertion in the main text and in Appendix \ref{app:params} that we are permitted to calculate the QMI at hand-selected times where $D(t)\approx 0$ for each parameter set separately. To support this, we show in Fig.~\ref{fig:QMI_revival} the QMI (for particle traces in the fermionic case in Set NE) $I_\text{part}^{\text{F}}$ as a function of the fraction of the environment, as in Fig.~\ref{fig:pQMI} (a), but at different times. In particular, we choose times close to one of the revivals in Set NE, where $D(t)\approx 0.8$. Clearly, whenever $D(t)$ substantially deviates from zero, the impurity is not decohered, and so the pointer basis we have been assuming is no longer a valid pointer basis. At a fully decohered time, we would have that $I_\text{part}^{\text{F}}[\text{I}:\hat D_{1\dots N_\text{E}}]\approx 2$ when the entire environment is included. But during a revival, we have  $I_\text{part}^{\text{F}}[\text{I}:D_{1\dots N_\text{E}}]< 2$. Ultimately this is explained by the fact that during a revival, $S(\hat \rho_\text{I})<1$ -- the impurity itself does not contain sufficient information in the position basis to pass to the environment. Overall, the QMI curve looks similar for all the time steps, with only the slope changing. But only when it is fully decohered do we have $I_\text{part}^{\text{F}}[\text{I}:\hat D_{1\dots N_\text{E}}]\approx 2$ anywhere. This means that, for a fair analysis of QD effects in the QMI, we must choose times when $D(t)\approx 0$.
To further bolster this claim, we show in Fig.~\ref{fig:QMI_SetB_time} (for fermionic Set S) how the QMI varies with time. Here, individual points in Fig.~\ref{fig:pQMI} (a) (representing the mutual information for the $p$-particle reduced density matrices) are entire curves in Fig.~\ref{fig:QMI_SetB_time} as functions of time. In this case, because there is heavy scrambling, once the system is equilibrated there are no revivals, meaning for instance that $I_\text{part}^{\text{F}}$ is almost exactly 2 for all post-equilibration when all of the environment particles are considered. But the other curves show more subtle effects.
Initially, when the dynamics are dominated by the interaction between the impurity and the environmental particles, the corresponding QMI curves indicate the appearance of a redundancy plateau. Since $|W_\text{IE}|>|W_\text{EE}|$ in Set S, the dynamics here are faster than the scrambling of information due to the environment-environment interaction. On later time scales, however, we observe that this `redundancy plateau' is lost and the QMI takes on the typical shape of a random state (see Fig.~\ref{fig:pQMI} (a)). This aligns with the expectation that the intra-environment interactions destroy any redundancy by scrambling the information \cite{blume05}. We find that the QMI reaches its equilibrium at $t\approx 25$, and only shows fluctuations after that time.

\section{System size analysis}\label{app:sys_size} 

\begin{figure*}[t]
    \centering
    \begin{minipage}[t]{0.48\textwidth}
        \centering
        \includegraphics[width=\linewidth]{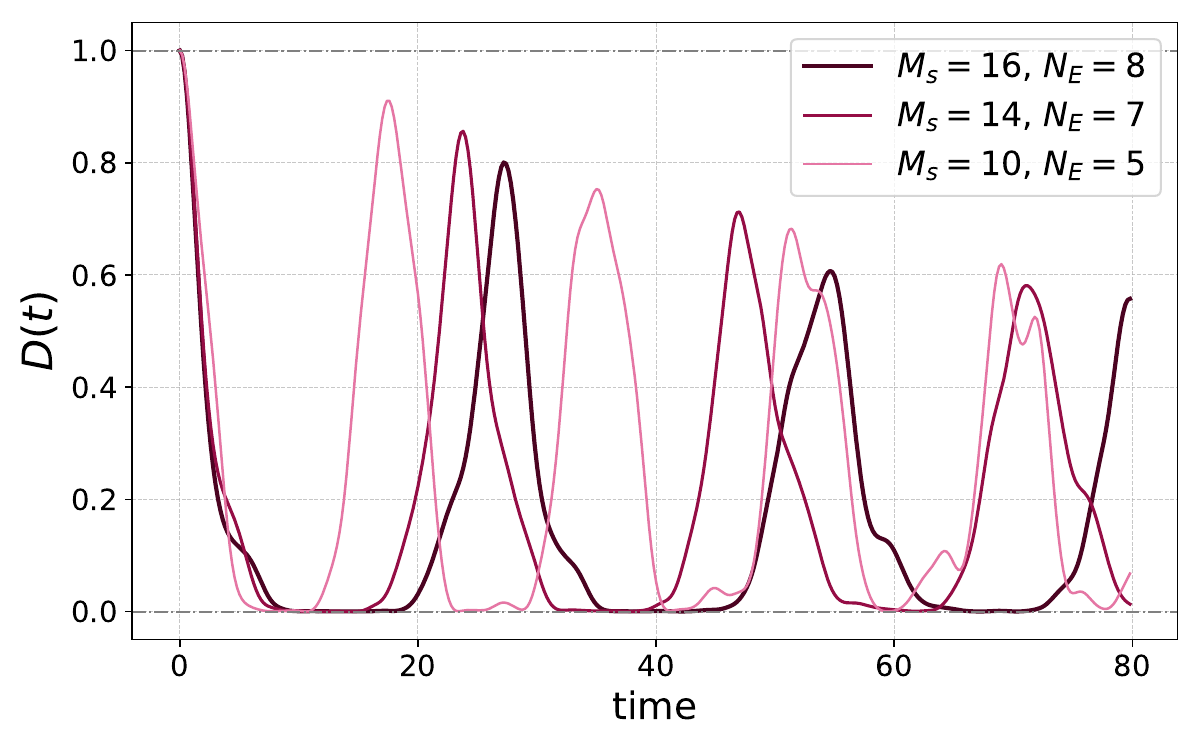}
        \caption{The decoherence parameter $D(t)$ propagated over time for different system sizes, for fermionic Set NE (which is known to have many large partial revivals). Comparing the revival times and the Hilbert space dimensions, we find the times and heights of the revivals decrease with increasing system size, which aligns with the macroscopic expectation of $D(t) \rightarrow 0$.}
        \label{fig:dist_dim}
    \end{minipage}
    \hfill
    \begin{minipage}[t]{0.48\textwidth}
        \centering
        \includegraphics[width=\linewidth]{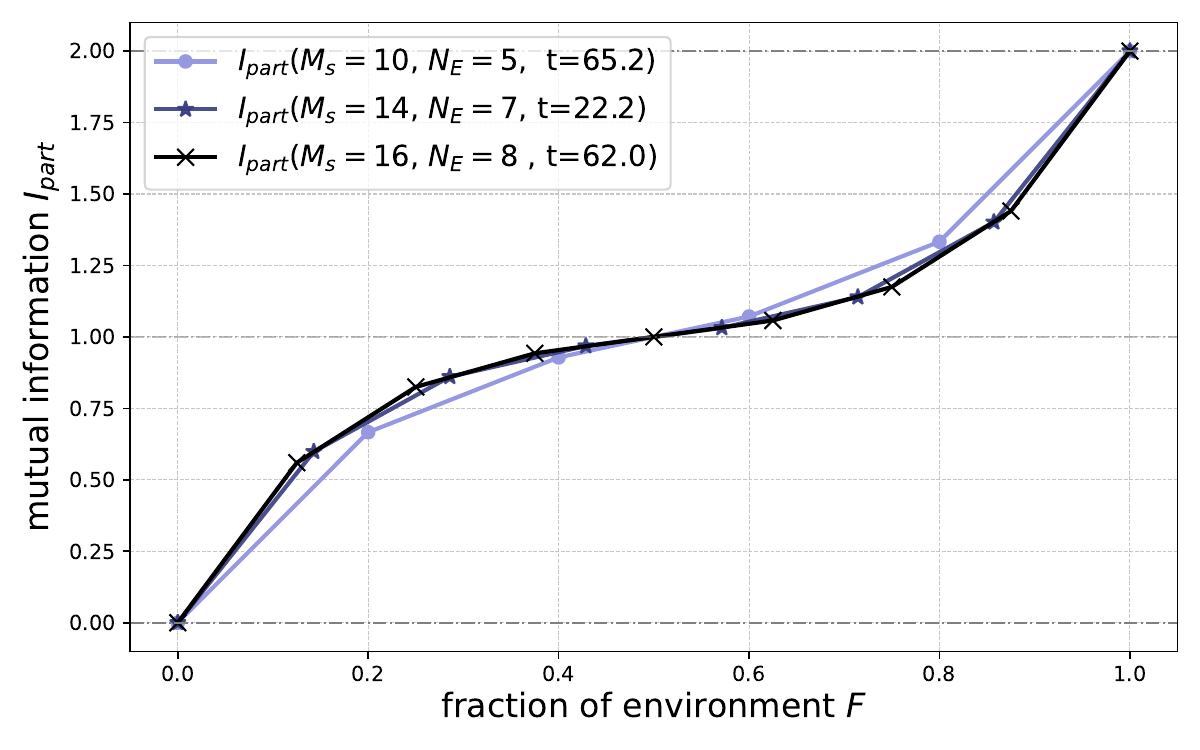}
        \caption{The QMI based on particle traces for fermionic Set R-LR, for increasing system sizes. The range of $i_\text{max}=M_\text{S}/2$ for all the cases.}
        \label{fig:MI_SetENC_dimH}
    \end{minipage}
\end{figure*}

In this final Appendix, we present how the observed trends from the main text change as the system size increases, to justify our assertion that the system is sufficiently large to study for many-body effects. First, in Fig.~\ref{fig:dist_dim} we show how the revivals in $D(t)$ that we see in Set NE change when the system size changes.

We observe that with increasing system size, the time spans over which we observe near perfect decoherence get longer. This is because the environmental particles take longer to be reflected back to the centre of the system. Then the size of the revivals in $D(t)$ decrease for increasing system size due to the larger number of states that are involved in the dynamics.
It is also interesting to investigate how the redundancy plateau that we observe (for instance for Set R-LR) changes with increasing system size. While the redundancy plateau flattens when increasing the system size from ($M_\text{S}=10$, $N_\text{E}=5$) to ($M_\text{S}=14$, $N_\text{E}=7$), in Fig.~\ref{fig:MI_SetENC_dimH} we see that for ($M_\text{S}=16$, $N_\text{E}=8$) the plateau seems to be practically saturated.
This is due to the fact that the environment particles cannot doubly-occupy a site. So despite the lack of a $W_{\text{EE}}$ in Set R and Set R-LR, the environment particles still influence each other by blocking each other from reaching the impurity -- they cannot fully gather around it. So even for increasing system size, there is a severe limitation to how much information a given additional environment particle can contain about the impurity location, and so the plateau can never be fully realised.
\bibliography{paper_refs}

@book{vonNeumann1930,
  author =        {Von Neumann, John},
  publisher =     {Princeton university press},
  title =         {Mathematical foundations of quantum mechanics: New
                   edition},
  year =          {2018},
  url =           {https://press.princeton.edu/books/hardcover/9780691178561/
                  mathematical-foundations-of-quantum-mechanics},
}

@Book{09WisemanMilburn,
  author    = {Wiseman, Howard M and Milburn, Gerard J},
  publisher = {Cambridge university press},
  title     = {Quantum measurement and control},
  year      = {2009},
  doi       = {10.1017/CBO9780511813948},
}

@Book{96BuschLahtiMittelstaedt,
  author    = {Busch, Paul and Lahti, Pekka J and Mittelstaedt, Peter},
  publisher = {Springer},
  title     = {The quantum theory of measurement},
  year      = {1996},
  doi       = {10.1007/978-3-540-37205-9},
}

@article{bell_against_measurement,
  author =        {Bell, John},
  journal =       {Physics world},
  number =        {8},
  pages =         {33--41},
  title =         {Against ‘measurement’},
  volume =        {3},
  year =          {1990},
  doi =           {10.1088/2058-7058/3/8/26},
}

@article{18FrauchigerRenner,
  author =        {Frauchiger, Daniela and Renner, Renato},
  journal =       {Nature communications},
  number =        {1},
  pages =         {3711},
  publisher =     {Nature Publishing Group UK London},
  title =         {Quantum theory cannot consistently describe the use
                   of itself},
  volume =        {9},
  year =          {2018},
  url =           {https://www.nature.com/articles/s41467-018-05739-8},
}

@InCollection{baumann20,
  author    = {Baumann, Veronika and Brukner, {\v{C}}aslav},
  booktitle = {Quantum, probability, logic: the work and influence of Itamar Pitowsky},
  publisher = {Springer},
  title     = {Wigner’s friend as a rational agent},
  year      = {2020},
  pages     = {91--99},
  doi       = {10.1007/978-3-030-34316-3_4},
}

@Article{relano20,
  author    = {Rela{\~n}o, Armando},
  journal   = {Physical Review A},
  title     = {Decoherence framework for Wigner's-friend experiments},
  year      = {2020},
  number    = {3},
  pages     = {032107},
  volume    = {101},
  doi       = {10.1103/PhysRevA.101.032107},
  publisher = {APS},
}

@article{25RivlinEngineerBaumann,
  author =        {Rivlin, Tom and Engineer, Sophie and
                   Baumann, Veronika},
  journal =       {arXiv preprint arXiv:2507.21221},
  title =         {Emergence of Classicality in Wigner's Friend
                   Scenarios},
  year =          {2025},
  url =           {https://arxiv.org/abs/2507.21221},
}

@Article{szilard1929,
  author    = {Szilard, Leo},
  journal   = {Zeitschrift f{\"u}r Physik},
  title     = {{\"U}ber die Entropieverminderung in einem thermodynamischen System bei Eingriffen intelligenter Wesen},
  year      = {1929},
  number    = {11},
  pages     = {840--856},
  volume    = {53},
  doi       = {10.1007/bf01341281},
  publisher = {Springer},
}

@Article{64Szilard,
  author    = {Szilard, Leo},
  journal   = {Behavioral Science},
  title     = {On the decrease of entropy in a thermodynamic system by the intervention of intelligent beings},
  year      = {1964},
  number    = {4},
  pages     = {301--310},
  volume    = {9},
  doi       = {10.1002/bs.3830090402},
  publisher = {Wiley Online Library},
}

@Article{peres80,
  author    = {Peres, Asher},
  journal   = {Physical Review D},
  title     = {Can we undo quantum measurements?},
  year      = {1980},
  number    = {4},
  pages     = {879},
  volume    = {22},
  doi       = {10.1103/PhysRevD.22.879},
  publisher = {APS},
}

@InCollection{83Zurek,
  author    = {Zurek, Wojciech Hubert},
  booktitle = {Quantum Optics, Experimental Gravity, and Measurement Theory},
  publisher = {Springer},
  title     = {Information transfer in quantum measurements: Irreversibility and amplification},
  year      = {1983},
  pages     = {87--116},
  doi       = {10.1007/978-1-4613-3712-6_5},
}

@Article{13Hormoz,
  author    = {Hormoz, Sahand},
  journal   = {Physical Review E—Statistical, Nonlinear, and Soft Matter Physics},
  title     = {Quantum collapse and the second law of thermodynamics},
  year      = {2013},
  number    = {2},
  pages     = {022129},
  volume    = {87},
  doi       = {10.1103/PhysRevE.87.022129},
  publisher = {APS},
}

@article{QThermoReview16,
  author =        {Goold, John and Huber, Marcus and Riera, Arnau and
                   Rio, L{\'\i}dia del and Skrzypczyk, Paul},
  journal =       {Journal of Physics A: Mathematical and Theoretical},
  number =        {14},
  pages =         {143001},
  publisher =     {IOP Publishing},
  title =         {The role of quantum information in thermodynamics—a
                   topical review},
  volume =        {49},
  year =          {2016},
  doi =           {10.1088/1751-8113/49/14/143001},
}

@Article{20GuryanovaFriisHuber,
  author    = {Guryanova, Yelena and Friis, Nicolai and Huber, Marcus},
  journal   = {Quantum},
  title     = {Ideal projective measurements have infinite resource costs},
  year      = {2020},
  pages     = {222},
  volume    = {4},
  doi       = {10.22331/q-2020-01-13-222},
  publisher = {Verein zur F{\"o}rderung des Open Access Publizierens in den Quantenwissenschaften},
}

@Article{carroll2021energy,
  author    = {Carroll, Sean M and Lodman, Jackie},
  journal   = {Foundations of Physics},
  title     = {Energy non-conservation in quantum mechanics},
  year      = {2021},
  number    = {4},
  pages     = {83},
  volume    = {51},
  doi       = {10.1007/s10701-021-00490-5},
  publisher = {Springer},
}

@Article{23MohammadyMiyadera,
  author    = {Mohammady, M Hamed and Miyadera, Takayuki},
  journal   = {Physical Review A},
  title     = {Quantum measurements constrained by the third law of thermodynamics},
  year      = {2023},
  number    = {2},
  pages     = {022406},
  volume    = {107},
  doi       = {10.1103/PhysRevA.107.022406},
  publisher = {APS},
}

@Article{17ElouardHerreraAuffeves,
  author    = {Elouard, Cyril and Herrera-Mart{\'\i}, David and Huard, Benjamin and Auffeves, Alexia},
  journal   = {Physical Review Letters},
  title     = {Extracting work from quantum measurement in Maxwell’s demon engines},
  year      = {2017},
  number    = {26},
  pages     = {260603},
  volume    = {118},
  doi       = {10.1103/PhysRevLett.118.260603},
  publisher = {APS},
}

@Article{23TarantoBakhshinezhadBluhm,
  author    = {Taranto, Philip and Bakhshinezhad, Faraj and Bluhm, Andreas and Silva, Ralph and Friis, Nicolai and Lock, Maximilian PE and Vitagliano, Giuseppe and Binder, Felix C and Debarba, Tiago and Schwarzhans, Emanuel and Huber, Marcus and Clivaz, Fabien},
  journal   = {PRX Quantum},
  title     = {Landauer Versus Nernst: What is the True Cost of Cooling a Quantum System?},
  year      = {2023},
  number    = {1},
  pages     = {010332},
  volume    = {4},
  doi       = {10.1103/PRXQuantum.4.010332},
  publisher = {APS},
}

@article{debarba2024broadcasting,
  author =        {Tiago Debarba and Marcus Huber and Nicolai Friis},
  journal =       {arXiv preprint arXiv:2403.07660},
  title =         {Unknown measurement statistics cannot be redundantly
                   copied using finite resources},
  year =          {2024},
  url =           {https://arxiv.org/abs/2403.07660},
}

@article{25MohammadyBuscemi,
  author =        {Mohammady, M Hamed and Buscemi, Francesco},
  journal =       {arXiv preprint arXiv:2502.14136},
  title =         {Thermodynamic closure of quantum measurements and the
                   limits of the indirect measurement model},
  year =          {2025},
  url =           {https://arxiv.org/abs/2502.14136},
}

@Article{candeloro26,
  author    = {Candeloro, Alessandro and Debarba, Tiago and Binder, Felix C},
  journal   = {Physical Review A},
  title     = {Thermodynamic constraints on the emergence of intersubjectivity in quantum systems},
  year      = {2026},
  number    = {3},
  pages     = {032201},
  volume    = {113},
  doi       = {10.1103/3312-b63x},
  publisher = {APS},
}

@Article{ballesteros26,
  author    = {Ballesteros Ferraz, Lorena and Elouard, Cyril},
  journal   = {Physical Review Research},
  title     = {Weak continuous measurements require more work than strong ones},
  year      = {2026},
  number    = {1},
  pages     = {013305},
  volume    = {8},
  doi       = {10.1103/6p2c-nbg5},
  publisher = {APS},
}

@Article{allahverdyan13,
  author    = {Allahverdyan, Armen E and Balian, Roger and Nieuwenhuizen, Theo M},
  journal   = {Physics Reports},
  title     = {Understanding quantum measurement from the solution of dynamical models},
  year      = {2013},
  number    = {1},
  pages     = {1--166},
  volume    = {525},
  doi       = {10.1016/j.physrep.2012.11.001},
  publisher = {Elsevier},
}

@article{23ArtiniPaternostro,
  author =        {Artini, Simone and Paternostro, Mauro},
  journal =       {New Journal of Physics},
  number =        {12},
  pages =         {123047},
  publisher =     {IOP Publishing},
  title =         {Characterizing the spontaneous collapse of a
                   wavefunction through entropy production},
  volume =        {25},
  year =          {2023},
  doi =           {10.1088/1367-2630/ad153a},
}

@Article{25ArtiniLoMonacoPaternostro,
  author    = {Artini, Simone and Lo Monaco, Gabriele and Donadi, Sandro and Paternostro, Mauro},
  journal   = {Physical Review Research},
  title     = {Nonequilibrium thermodynamics of gravitational objective-collapse models},
  year      = {2025},
  number    = {4},
  pages     = {043017},
  volume    = {7},
  doi       = {10.1103/7hqd-zf96},
  publisher = {APS},
}

@Article{25WallsBlossFord,
  author    = {Walls, Sophia M. and Bloss, Adam and Ford, Ian J.},
  journal   = {Phys. Rev. A},
  title     = {Characterizing quantum measurement through environmental stochastic entropy production in a two-spin-1/2 system},
  year      = {2025},
  month     = sep,
  pages     = {032210},
  volume    = {112},
  doi       = {10.1103/jdwl-rv38},
  publisher = {American Physical Society},
}

@Article{25LatuneElouard,
  author    = {Latune, Camille L and Elouard, Cyril},
  journal   = {Quantum},
  title     = {A thermodynamically consistent approach to the energy costs of quantum measurements},
  year      = {2025},
  pages     = {1614},
  volume    = {9},
  doi       = {10.22331/q-2025-01-28-1614},
  publisher = {Verein zur F{\"o}rderung des Open Access Publizierens in den Quantenwissenschaften},
}

@Article{schwarzhans25,
  author    = {Schwarzhans, Emanuel and Binder, Felix C and Huber, Marcus and Lock, Maximilian PE},
  journal   = {Physical Review Research},
  title     = {Quantum measurements and equilibration: The emergence of objective outcomes via entropy maximization},
  year      = {2025},
  number    = {4},
  pages     = {043279},
  volume    = {7},
  doi       = {10.1103/h2yj-25rn},
  publisher = {APS},
}

@Article{schwarzhans26,
  author    = {Schwarzhans, Emanuel and Apollaro, Tony JG and Khomchenko, Ilia and Lock, Maximilian PE and Mitchison, Mark T and Huber, Marcus},
  journal   = {PRX Quantum},
  title     = {Quantum detectors as autonomous machines: assessing the nonequilibrium thermodynamics of information acquisition},
  year      = {2026},
  number    = {3},
  pages     = {033001},
  volume    = {7},
  doi       = {10.1103/wm5p-tjtg},
  publisher = {APS},
}

@Article{engineer26,
  author    = {Engineer, Sophie and Rivlin, Tom and Wollmann, Sabine and Malik, Mehul and Lock, Maximilian PE},
  journal   = {Physical Review A},
  title     = {Equilibration of objective observables in a dynamical model of quantum measurements},
  year      = {2026},
  number    = {3},
  pages     = {032205},
  volume    = {113},
  doi       = {10.1103/lkqb-jdmg},
  publisher = {APS},
}

@Article{zurek_1981_pointer,
  author    = {Zurek, W. H.},
  journal   = {Phys. Rev. D},
  title     = {Pointer basis of quantum apparatus: Into what mixture does the wave packet collapse?},
  year      = {1981},
  month     = sep,
  pages     = {1516--1525},
  volume    = {24},
  doi       = {10.1103/PhysRevD.24.1516},
  publisher = {American Physical Society},
}

@Article{zurek_1982_environment,
  author    = {Zurek, W. H.},
  journal   = {Phys. Rev. D},
  title     = {Environment-induced superselection rules},
  year      = {1982},
  month     = oct,
  pages     = {1862--1880},
  volume    = {26},
  doi       = {10.1103/PhysRevD.26.1862},
  publisher = {American Physical Society},
}

@Article{joos_emergence_1985,
  author  = {Joos, E. and Zeh, H. D.},
  journal = {Zeitschrift für Physik B Condensed Matter},
  title   = {The emergence of classical properties through interaction with the environment},
  year    = {1985},
  issn    = {0722-3277, 1434-6036},
  month   = jun,
  number  = {2},
  pages   = {223--243},
  volume  = {59},
  doi     = {10.1007/BF01725541},
}

@Article{schlosshauer_2005_decoherence,
  author    = {Schlosshauer, Maximilian},
  journal   = {Rev. Mod. Phys.},
  title     = {Decoherence, the measurement problem, and interpretations of quantum mechanics},
  year      = {2005},
  month     = feb,
  pages     = {1267--1305},
  volume    = {76},
  doi       = {10.1103/RevModPhys.76.1267},
  publisher = {American Physical Society},
}

@Article{22Zurek,
  author    = {Zurek, Wojciech Hubert},
  journal   = {Entropy},
  title     = {Quantum Theory of the Classical: Einselection, Envariance, Quantum Darwinism and Extantons},
  year      = {2022},
  number    = {11},
  pages     = {1520},
  volume    = {24},
  doi       = {10.3390/e24111520},
  publisher = {MDPI},
}

@Article{luders_uber_1950,
  author  = {Lüders, Gerhart},
  journal = {Annalen der Physik},
  title   = {Über die {Zustandsänderung} durch den {Me\ss proze\ss }},
  year    = {1950},
  issn    = {0003-3804, 1521-3889},
  month   = jan,
  number  = {5-8},
  pages   = {322--328},
  volume  = {443},
  doi     = {10.1002/andp.19504430510},
}

@article{09Zurek,
  author =        {Zurek, Wojciech Hubert},
  journal =       {Nature physics},
  number =        {3},
  pages =         {181--188},
  publisher =     {Nature Publishing Group UK London},
  title =         {Quantum Darwinism},
  volume =        {5},
  year =          {2009},
  url =           {https://www.nature.com/articles/nphys1202},
}

@Article{15HorodeckiKorbiczHorodecki,
  author    = {Horodecki, Ryszard and Korbicz, JK and Horodecki, Pawe{\l}},
  journal   = {Physical review A},
  title     = {Quantum origins of objectivity},
  year      = {2015},
  number    = {3},
  pages     = {032122},
  volume    = {91},
  doi       = {10.1103/PhysRevA.91.032122},
  publisher = {APS},
}

@Article{21Korbicz,
  author    = {Korbicz, JK},
  journal   = {Quantum},
  title     = {Roads to objectivity: quantum darwinism, spectrum broadcast structures, and strong quantum darwinism--a review},
  year      = {2021},
  pages     = {571},
  volume    = {5},
  doi       = {10.22331/q-2021-11-08-571},
  publisher = {Verein zur F{\"o}rderung des Open Access Publizierens in den Quantenwissenschaften},
}

@Article{ollivier_objective_2004,
  author  = {Ollivier, Harold and Poulin, David and Zurek, Wojciech H.},
  journal = {Physical Review Letters},
  title   = {Objective {Properties} from {Subjective} {Quantum} {States}: {Environment} as a {Witness}},
  year    = {2004},
  issn    = {0031-9007, 1079-7114},
  month   = nov,
  number  = {22},
  pages   = {220401},
  volume  = {93},
  doi     = {10.1103/PhysRevLett.93.220401},
}

@Article{blume05,
  author    = {Blume-Kohout, Robin and Zurek, Wojciech H},
  journal   = {Foundations of Physics},
  title     = {A simple example of “Quantum Darwinism”: Redundant information storage in many-spin environments},
  year      = {2005},
  number    = {11},
  pages     = {1857--1876},
  volume    = {35},
  doi       = {10.1007/s10701-005-7352-5},
  publisher = {Springer},
}

@article{riedel12,
  author =        {Riedel, C Jess and Zurek, Wojciech H and
                   Zwolak, Michael},
  journal =       {New Journal of physics},
  number =        {8},
  pages =         {083010},
  publisher =     {IOP Publishing},
  title =         {The rise and fall of redundancy in decoherence and
                   quantum Darwinism},
  volume =        {14},
  year =          {2012},
  doi =           {10.1088/1367-2630/14/8/083010},
}

@article{galve2016non,
  author =        {Galve, Fernando and Zambrini, Roberta and
                   Maniscalco, Sabrina},
  journal =       {Scientific reports},
  number =        {1},
  pages =         {19607},
  publisher =     {Nature Publishing Group UK London},
  title =         {Non-markovianity hinders quantum darwinism},
  volume =        {6},
  year =          {2016},
  url =           {https://www.nature.com/articles/srep19607},
}

@Article{17MironowiczKorbiczHorodecki,
  author    = {Mironowicz, Piotr and Korbicz, JK and Horodecki, Pawe{\l}},
  journal   = {Physical review letters},
  title     = {Monitoring of the process of system information broadcasting in time},
  year      = {2017},
  number    = {15},
  pages     = {150501},
  volume    = {118},
  doi       = {10.1103/PhysRevLett.118.150501},
  publisher = {APS},
}

@Article{19LeOlayaCastro,
  author    = {Le, Thao P and Olaya-Castro, Alexandra},
  journal   = {Physical review letters},
  title     = {Strong quantum darwinism and strong independence are equivalent to spectrum broadcast structure},
  year      = {2019},
  number    = {1},
  pages     = {010403},
  volume    = {122},
  doi       = {10.1103/PhysRevLett.122.010403},
  publisher = {APS},
}

@Article{20LeOlayaCastro,
  author    = {Le, Thao P and Olaya-Castro, Alexandra},
  journal   = {Quantum Science and Technology},
  title     = {Witnessing non-objectivity in the framework of strong quantum Darwinism},
  year      = {2020},
  month     = aug,
  number    = {4},
  pages     = {045012},
  volume    = {5},
  doi       = {10.1088/2058-9565/abac4e},
  publisher = {IOP Publishing},
}

@Article{22TouilYanGirolami,
  author    = {Touil, Akram and Yan, Bin and Girolami, Davide and Deffner, Sebastian and Zurek, Wojciech Hubert},
  journal   = {Physical review letters},
  title     = {Eavesdropping on the decohering environment: Quantum darwinism, amplification, and the origin of objective classical reality},
  year      = {2022},
  number    = {1},
  pages     = {010401},
  volume    = {128},
  doi       = {10.1103/PhysRevLett.128.010401},
  publisher = {APS},
}

@Article{duruisseau23,
  author    = {Duruisseau, Paul and Touil, Akram and Deffner, Sebastian},
  journal   = {Entropy},
  title     = {Pointer states and quantum Darwinism with two-body interactions},
  year      = {2023},
  number    = {12},
  pages     = {1573},
  volume    = {25},
  doi       = {10.3390/e25121573},
  publisher = {Mdpi},
}

@Article{chisholm23,
  author    = {Chisholm, Diana A and Innocenti, Luca and Palma, G Massimo},
  journal   = {Quantum},
  title     = {The meaning of redundancy and consensus in quantum objectivity},
  year      = {2023},
  pages     = {1074},
  volume    = {7},
  doi       = {10.22331/q-2023-08-03-1074},
  publisher = {Verein zur F{\"o}rderung des Open Access Publizierens in den Quantenwissenschaften},
}

@Article{doucet24,
  author    = {Doucet, Emery and Deffner, Sebastian},
  journal   = {Physical Review X},
  title     = {Classifying Two-Body Hamiltonians for Quantum Darwinism},
  year      = {2024},
  number    = {4},
  pages     = {041064},
  volume    = {14},
  doi       = {10.1103/PhysRevX.14.041064},
  publisher = {APS},
}

@Article{chisholm24,
  author    = {Chisholm, Diana A and Innocenti, Luca and Palma, G Massimo},
  journal   = {Physical Review A},
  title     = {Importance of using the averaged mutual information when quantifying quantum objectivity},
  year      = {2024},
  number    = {1},
  pages     = {012218},
  volume    = {110},
  doi       = {10.1103/PhysRevA.110.012218},
  publisher = {APS},
}

@Article{chisholm_emergence_2026,
  author  = {Chisholm, Diana A. and Palma, G. Massimo and Innocenti, Luca},
  journal = {APS Open Science},
  title   = {Emergence of quantum {Darwinism} and pointer states for noncommuting evolutions},
  year    = {2026},
  issn    = {3070-2240},
  month   = may,
  pages   = {000024},
  volume  = {1},
  doi     = {10.1103/75jr-ltct},
}

@Article{kiely26,
  author    = {Kiely, Anthony and Chisholm, Diana A and Touil, Akram and Deffner, Sebastian and Landi, Gabriel and Campbell, Steve},
  journal   = {Physical Review A},
  title     = {Metrological approach to the emergence of classical objectivity},
  year      = {2026},
  number    = {2},
  pages     = {022403},
  volume    = {113},
  doi       = {10.1103/hn78-7xx3},
  publisher = {APS},
}

@Article{18CiampiniPinnaPaternostro,
  author    = {Ciampini, Mario A. and Pinna, Giorgia and Mataloni, Paolo and Paternostro, Mauro},
  journal   = {Physical Review A},
  title     = {Experimental signature of quantum Darwinism in photonic cluster states},
  year      = {2018},
  issn      = {2469-9934},
  month     = aug,
  number    = {2},
  volume    = {98},
  doi       = {10.1103/physreva.98.020101},
  publisher = {American Physical Society (APS)},
}

@Article{25ZhuSaliceTouil,
  author    = {Zhu, Zitian and Salice, Kiera and Touil, Akram and Bao, Zehang and Song, Zixuan and Zhang, Pengfei and Li, Hekang and Wang, Zhen and Song, Chao and Guo, Qiujiang and others},
  journal   = {Science Advances},
  title     = {Observation of quantum Darwinism and the origin of classicality with superconducting circuits},
  year      = {2025},
  number    = {31},
  pages     = {eadx6857},
  volume    = {11},
  doi       = {10.1126/sciadv.adx6857},
  publisher = {American Association for the Advancement of Science},
}

@Article{15GiorgiGalveZambrini,
  author    = {Giorgi, Gian Luca and Galve, Fernando and Zambrini, Roberta},
  journal   = {Physical Review A},
  title     = {Quantum Darwinism and non-Markovian dissipative dynamics from quantum phases of the spin-1/2 XX model},
  year      = {2015},
  number    = {2},
  pages     = {022105},
  volume    = {92},
  doi       = {10.1103/PhysRevA.92.022105},
  publisher = {APS},
}

@article{16ZwolakRiedelZurek,
  author =        {Zwolak, Michael and Riedel, C Jess and
                   Zurek, Wojciech H},
  journal =       {Scientific Reports},
  number =        {1},
  pages =         {25277},
  publisher =     {Nature Publishing Group UK London},
  title =         {Amplification, decoherence and the acquisition of
                   information by spin environments},
  volume =        {6},
  year =          {2016},
  url =           {https://www.nature.com/articles/srep25277},
}

@article{21MirkinWisniacki,
  author =        {Mirkin, Nicol{\'a}s and Wisniacki, Diego A},
  journal =       {Entropy},
  number =        {11},
  pages =         {1377},
  publisher =     {MDPI},
  title =         {Many-body localization and the emergence of quantum
                   Darwinism},
  volume =        {23},
  year =          {2021},
  doi =           {10.3390/e23111377},
}

@Article{21RyanPaternostroCampbell,
  author    = {Ryan, Eoghan and Paternostro, Mauro and Campbell, Steve},
  journal   = {Physics Letters A},
  title     = {Quantum Darwinism in a structured spin environment},
  year      = {2021},
  pages     = {127675},
  volume    = {416},
  doi       = {10.1016/j.physleta.2021.127675},
  publisher = {Elsevier},
}

@article{21KwiatkowskiCywinskiKorbicz,
  author =        {Kwiatkowski, Damian and Cywi{\'n}ski, {\L}ukasz and
                   Korbicz, Jaros{\l}aw K},
  journal =       {New Journal of Physics},
  number =        {4},
  pages =         {043036},
  publisher =     {IOP Publishing},
  title =         {Appearance of objectivity for NV centers interacting
                   with dynamically polarized nuclear environment},
  volume =        {23},
  year =          {2021},
  doi =           {10.1088/1367-2630/abeffd},
}

@article{bloch05,
  author =        {Bloch, Immanuel},
  journal =       {Nature physics},
  number =        {1},
  pages =         {23--30},
  publisher =     {Nature Publishing Group UK London},
  title =         {Ultracold quantum gases in optical lattices},
  volume =        {1},
  year =          {2005},
  url =           {https://www.nature.com/articles/nphys138},
}

@article{bloch_quantum_2012,
  author =        {Bloch, Immanuel and Dalibard, Jean and
                   Nascimbène, Sylvain},
  journal =       {Nature Physics},
  month =         apr,
  number =        {4},
  pages =         {267--276},
  title =         {Quantum simulations with ultracold quantum gases},
  volume =        {8},
  year =          {2012},
  doi =           {10.1038/nphys2259},
  issn =          {1745-2473, 1745-2481},
  url =           {https://www.nature.com/articles/nphys2259},
}

@Article{gross_quantum_2017,
  author  = {Gross, Christian and Bloch, Immanuel},
  journal = {Science},
  title   = {Quantum simulations with ultracold atoms in optical lattices},
  year    = {2017},
  issn    = {0036-8075, 1095-9203},
  month   = sep,
  number  = {6355},
  pages   = {995--1001},
  volume  = {357},
  doi     = {10.1126/science.aal3837},
}

@article{26KendrickKaleGreiner,
  author =        {Kendrick, Lev Haldar and Kale, Anant and Gang, Youqi and
                   Deters, Alexander Dennisovich and Lebrat, Martin and
                   Young, Aaron W. and Greiner, Markus},
  journal =       {Nature Physics},
  title =         {Pseudogap in a Fermi–Hubbard quantum simulator},
  year =          {2026},
  doi =           {10.1038/s41586-026-10875-z},
  issn =          {0028-0836, 1476-4687},
  url =           {https://www.nature.com/articles/s41586-026-10875-z},
}

@article{rigol08,
  author =        {Rigol, Marcos and Dunjko, Vanja and Olshanii, Maxim},
  journal =       {Nature},
  number =        {7189},
  pages =         {854--858},
  publisher =     {Nature Publishing Group UK London},
  title =         {Thermalization and its mechanism for generic isolated
                   quantum systems},
  volume =        {452},
  year =          {2008},
  url =           {https://www.nature.com/articles/nature06838},
}

@article{16GogolinEisert,
  author =        {Gogolin, Christian and Eisert, Jens},
  journal =       {Reports on Progress in Physics},
  number =        {5},
  pages =         {056001},
  publisher =     {IOP Publishing},
  title =         {Equilibration, thermalisation, and the emergence of
                   statistical mechanics in closed quantum systems},
  volume =        {79},
  year =          {2016},
  doi =           {10.1088/0034-4885/79/5/056001},
}

@Article{Meier25,
  author    = {Meier, Florian and Rivlin, Tom and Debarba, Tiago and Xuereb, Jake and Huber, Marcus and Lock, Maximilian PE},
  journal   = {PRX Quantum},
  title     = {Emergence of a second law of thermodynamics in isolated quantum systems},
  year      = {2025},
  number    = {1},
  pages     = {010309},
  volume    = {6},
  doi       = {10.1103/PRXQuantum.6.010309},
  publisher = {APS},
}

@Article{dalessio2016,
  author    = {D'Alessio, Luca and Kafri, Yariv and Polkovnikov, Anatoli and Rigol, Marcos},
  journal   = {Advances in Physics},
  title     = {From quantum chaos and eigenstate thermalization to statistical mechanics and thermodynamics},
  year      = {2016},
  number    = {3},
  pages     = {239--362},
  volume    = {65},
  doi       = {10.1080/00018732.2016.1198134},
  publisher = {Taylor \& Francis},
}

@article{deutsch2018eigenstate,
  author =        {Deutsch, Joshua M},
  journal =       {Reports on Progress in Physics},
  number =        {8},
  pages =         {082001},
  publisher =     {IOP Publishing},
  title =         {Eigenstate thermalization hypothesis},
  volume =        {81},
  year =          {2018},
  doi =           {10.1088/1361-6633/aac9f1},
}

@article{Cao26,
  author =        {Cao, Xiangyu and Nussinov, Zohar},
  journal =       {arXiv preprint arXiv:2603.15743v1},
  title =         {Redundancy from Subsystem Thermalization},
  year =          {2026},
  url =           {https://arxiv.org/abs/2603.15743},
}

@article{kuhr_quantum-gas_2016,
  author =        {Kuhr, Stefan},
  journal =       {National Science Review},
  month =         jun,
  number =        {2},
  pages =         {170--172},
  title =         {Quantum-gas microscopes: a new tool for cold-atom
                   quantum simulators},
  volume =        {3},
  year =          {2016},
  doi =           {10.1093/nsr/nww023},
  issn =          {2053-714X, 2095-5138},
  url =           {https://academic.oup.com/nsr/article/3/2/170/2460374},
}

@Article{10Esslinger,
  author    = {Esslinger, Tilman},
  journal   = {Annu. Rev. Condens. Matter Phys.},
  title     = {Fermi-Hubbard physics with atoms in an optical lattice},
  year      = {2010},
  number    = {1},
  pages     = {129--152},
  volume    = {1},
  doi       = {10.1146/annurev-conmatphys-070909-104059},
  publisher = {Annual Reviews},
}

@Article{22KourehpazDonsaBrezinova,
  author    = {Mahdi Kourehpaz and Stefan Donsa and Fabian Lackner and Joachim Burgd\"{o}rfer and Iva B{\v{r}}ezinov{\'{a}}},
  journal   = {Entropy},
  title     = {Canonical Density Matrices from Eigenstates of Mixed Systems},
  year      = {2022},
  month     = nov,
  number    = {12},
  pages     = {1740},
  volume    = {24},
  doi       = {10.3390/e24121740},
  publisher = {{MDPI} {AG}},
}

@Article{chomaz_dipolar_2023,
  author  = {Chomaz, Lauriane and Ferrier-Barbut, Igor and Ferlaino, Francesca and Laburthe-Tolra, Bruno and Lev, Benjamin L and Pfau, Tilman},
  journal = {Reports on Progress in Physics},
  title   = {Dipolar physics: a review of experiments with magnetic quantum gases},
  year    = {2023},
  issn    = {0034-4885, 1361-6633},
  month   = feb,
  number  = {2},
  pages   = {026401},
  volume  = {86},
  doi     = {10.1088/1361-6633/aca814},
}

@Book{18NegeleOrland,
  author    = {Negele, John W and Orland, Henri},
  publisher = {CRC Press},
  title     = {Quantum many-particle systems},
  year      = {2018},
  doi       = {10.1201/9780429497926},
}

@Article{blume2006quantum,
  author    = {Blume-Kohout, Robin and Zurek, Wojciech H},
  journal   = {Physical Review A},
  title     = {Quantum Darwinism: Entanglement, branches, and the emergent classicality of redundantly stored quantum information},
  year      = {2006},
  number    = {6},
  pages     = {062310},
  volume    = {73},
  doi       = {10.1103/PhysRevA.73.062310},
  publisher = {APS},
}

@article{touil2024branching,
  author =        {Touil, Akram and Anza, Fabio and Deffner, Sebastian and
                   Crutchfield, James P},
  journal =       {Quantum},
  pages =         {1494},
  publisher =     {Verein zur F{\"o}rderung des Open Access Publizierens
                   in den Quantenwissenschaften},
  title =         {Branching states as the emergent structure of a
                   quantum universe},
  volume =        {8},
  year =          {2024},
  doi =           {10.22331/q-2024-10-10-1494},
}

@Article{11Short,
  author    = {Short, Anthony J},
  journal   = {New Journal of Physics},
  title     = {Equilibration of quantum systems and subsystems},
  year      = {2011},
  number    = {5},
  pages     = {053009},
  volume    = {13},
  doi       = {10.1088/1367-2630/13/5/053009},
  publisher = {IOP Publishing},
}

@Book{nielsen_chuang_2010,
  author    = {Nielsen, Michael A. and Chuang, Isaac L.},
  publisher = {Cambridge University Press},
  title     = {Quantum Computation and Quantum Information: 10th Anniversary Edition},
  year      = {2010},
  doi       = {10.1017/CBO9780511976667},
}

@Article{ding_concept_2021,
  author    = {Ding, Lexin and Mardazad, Sam and Das, Sreetama and Szalay, Szilárd and Schollwöck, Ulrich and Zimborás, Zoltán and Schilling, Christian},
  journal   = {Journal of Chemical Theory and Computation},
  title     = {Concept of {Orbital} {Entanglement} and {Correlation} in {Quantum} {Chemistry}},
  year      = {2021},
  issn      = {1549-9618},
  month     = jan,
  number    = {1},
  pages     = {79--95},
  volume    = {17},
  doi       = {10.1021/acs.jctc.0c00559},
  publisher = {American Chemical Society},
}

@Article{coleman_structure_1963,
  author    = {Coleman, A. J.},
  journal   = {Rev. Mod. Phys.},
  title     = {Structure of Fermion Density Matrices},
  year      = {1963},
  month     = jul,
  pages     = {668--686},
  volume    = {35},
  doi       = {10.1103/RevModPhys.35.668},
  publisher = {American Physical Society},
}

@Article{amosov_spectral_2017,
  author  = {Amosov, Grigori G. and Filippov, Sergey N.},
  journal = {Quantum Information Processing},
  title   = {Spectral properties of reduced fermionic density operators and parity superselection rule},
  year    = {2017},
  issn    = {1570-0755, 1573-1332},
  month   = jan,
  number  = {1},
  pages   = {2},
  volume  = {16},
  doi     = {10.1007/s11128-016-1467-9},
}

@Article{galler_orbital_2021,
  author  = {Galler, Anna and Thunström, Patrik},
  journal = {Physical Review Research},
  title   = {Orbital and electronic entanglement in quantum teleportation schemes},
  year    = {2021},
  issn    = {2643-1564},
  month   = aug,
  number  = {3},
  pages   = {033120},
  volume  = {3},
  doi     = {10.1103/PhysRevResearch.3.033120},
}

@Article{ernst_mode_2024,
  author  = {Ernst, Jan Ole and Tennie, Felix},
  journal = {New Journal of Physics},
  title   = {Mode entanglement in fermionic and bosonic {Harmonium}},
  year    = {2024},
  issn    = {1367-2630},
  month   = mar,
  number  = {3},
  pages   = {033042},
  volume  = {26},
  doi     = {10.1088/1367-2630/ad240f},
}

@Article{deffner17,
  author    = {Deffner, Sebastian and Campbell, Steve},
  journal   = {Journal of Physics A: Mathematical and Theoretical},
  title     = {Quantum speed limits: from Heisenberg’s uncertainty principle to optimal quantum control},
  year      = {2017},
  number    = {45},
  pages     = {453001},
  volume    = {50},
  doi       = {10.1088/1751-8121/aa86c6/meta},
  publisher = {IOP Publishing},
}

@Article{07SenBrussLewenstein,
  author    = {Sen, Aditi and Sen, Ujjwal and Gromek, Bartosz and Bru{\ss}, Dagmar and Lewenstein, Maciej},
  journal   = {Physical Review A},
  title     = {Capacities of noiseless quantum channels for massive indistinguishable particles: Bosons versus fermions},
  year      = {2007},
  number    = {2},
  pages     = {022331},
  volume    = {75},
  doi       = {10.1103/PhysRevA.75.022331},
  publisher = {APS},
}

@article{09HaqueZozulyaSchoutens,
  author =        {Haque, Masudul and Zozulya, OS and
                   Schoutens, Kareljan},
  journal =       {Journal of Physics A: Mathematical and Theoretical},
  number =        {50},
  pages =         {504012},
  title =         {Entanglement between particle partitions in itinerant
                   many-particle states},
  volume =        {42},
  year =          {2009},
  doi =           {10.1088/1751-8113/42/50/504012},
}

@Article{26TengXuYang,
  author    = {Teng, Xiao-Wei and Xu, Rui-Yang and Yang, Hui-Chen and Wu, Shu-Min},
  journal   = {Nuclear Physics B},
  title     = {Bosonic and fermionic mutual information of N-partite systems in dilaton black hole background},
  year      = {2026},
  pages     = {117559},
  doi       = {10.1016/j.nuclphysb.2026.117559},
  publisher = {North-Holland},
}

\end{document}